\documentclass[fleqn]{elsart}
\journal{Annals of Physics}
\usepackage{graphicx}
\usepackage{dcolumn}
\usepackage{bm}
\usepackage{amssymb}
\usepackage{amsmath}
\usepackage{latexsym}
\newcommand{\be}{\begin{equation}}
\newcommand{\ee}{\end{equation}}
\newcommand{\bea}{\begin{eqnarray}}
\newcommand{\nn}{\nonumber}
\newcommand{\eea}{\end{eqnarray}}
\newcommand{\s}{\stackrel}

\begin{document}

\begin{frontmatter}

\title{Metric-affine $f(R)$ theories of gravity}
\author{Thomas P.~Sotiriou},
\ead{sotiriou@sissa.it}
\author{Stefano Liberati}
\ead{liberati@sissa.it}
\address{SISSA-ISAS, via Beirut 2-4, 34014, Trieste, Italy and INFN, Sezione di Trieste}
\begin{abstract}
 General Relativity assumes that spacetime is fully described by the metric alone. An alternative is the so called Palatini formalism where the metric and the connections are taken as independent quantities. The metric-affine theory of gravity has attracted considerable attention recently, since it was shown that within this framework some cosmological models, based on some generalized gravitational actions, can account for the current accelerated expansion of the universe. However we think that metric-affine gravity  deserves much more attention than that related to cosmological applications and so we consider here metric-affine gravity theories in which the gravitational action is a general function of the scalar curvature while the matter action is allowed to depend also on the connection which is not {\em a priori} symmetric. This general treatment will allow us to address several open issues such as: the relation between metric-affine $f(R)$ gravity and General Relativity (in vacuum as well as in the presence of matter), the implications of the dependence (or independence) of the matter action on the connections, 
the origin and role of torsion and
the viability of the minimal-coupling principle.
\end{abstract}

\end{frontmatter}

\section{Introduction}
\label{sec:intro}

 General relativity (GR) is certainly one of the most elegant scientific theories ever developed. Even more remarkably for almost a century this conceptual elegance went together with a notable success in explaining observations which was equaled only by the other pillar of modern physics, {\em i.e.}~quantum mechanics. Of course the scientific community has long been well aware of the intrinsic limits of GR which are revealed by the existence of solutions with curvature singularities or closed time-like loops. These were all considered hints that the theory had to be replaced, in some very strong field limit, by a more complete one, generally identified with a quantum theory of gravity. The latter has proved to be much more difficult to realize than expected, considering the fact that  we still do not have a complete theory of quantum gravity (QG) after several decades of research in this direction.

It is interesting to note that many attempts to achieve such quantization of gravity have shown that somehow a modification of the standard gravitational action (the Einstein--Hilbert one) seems to be necessary. In particular one can cite early results (see {\em e.g.}~Utiyama and De Witt~\cite{uti}) showing that renormalizability at one-loop demands that the classical action should be supplemented by higher order curvature terms. More recently there have been many calculations showing that, when quantum corrections or String/M-theory are considered, the effective classical gravitational action admits higher order corrections in curvature invariants~\cite{quant1,quant2,quant3,quant4,noji2,vassi}. 
However such results have often been considered evidence of the need for a modified theory whose relevance was going to be limited to very strong gravity phenomena ({\em e.g.}~early universe or black hole physics).

This framework has been radically changed in the last two decades mainly due to the growing observational evidence indicating that standard GR seems unable to explain crucial features of the current, low energy universe, without introducing some very artificial assumptions. In particular the dynamics of galaxies and clusters of galaxies as well as the present accelerated expansion of the universe seem to require the introduction of exotic, not yet directly observed, contributions to the matter content of the cosmos which go under the name of dark matter and dark energy. The observation that such ``dark'' components amount to about 96\% of our universe is at the  same time remarkable and frustrating \cite{Carroll:2001xs}. Even if dark matter could be accounted for by some stable super-symmetric particle yet to be detected, the dark energy component  would still make up about 70\% of our universe. Moreover, the simplest explanation for dark energy which is based on a cosmological constant, has both a problem of magnitude (the particle physics natural energy of the vacuum should be much higher than what we observe) and of timing (the so called coincidence problem: why is the cosmological constant starting to  dominate just now?)~\cite{Carroll:2000fy}. 
Of course it may well be that such observational evidence could find an explanation within the realm of the current available theories of particles 
but there seems to be rather compelling motivation for investigating whether there might be a viable alternative theory of gravity, which would avoid the necessity for introducing such exotic constituents.

Could it be that our inability to form an acceptable theory of quantum gravity as well as to get a simple explanation of the low energy cosmological observations --- which are probing scales at which gravity should behave essentially classically --- is due to using an over-restricted classical theory of gravity? The facts mentioned above seem to point in this direction. Therefore, we believe that serious attention should be paid to the classical formulation of gravity.

\subsection{Gravity beyond General Relativity}
\label{sec:beygr}

 Attempts to generalize GR have  started from the early years of the theory. Scalar tensor theories \cite{faraoni} and the Einstein--Cartan formulation~\cite{EC-RMP} are just two examples of a very broad variety of models. In the 1970s many of these alternative theories where severely constrained using the post-Newtonian formalism (see e.g~\cite{will,will2}). On the other hand there are several alternative formulations of gravity which are relevant only in regimes which are yet to be probed with the necessary precision. Examples in this sense are models based on the presence of extra dimensions, some of which are known as brane-world scenarios \cite{dvali,randal,marteens} 
or theories with actions involving extra positive powers of the scalar curvature.

Indeed the form of the gravitational action has been among the most questioned features of GR, as the Einstein--Hilbert action 
\begin{equation}
S_{EH}=\frac{1}{2\kappa}\int d^4 x \sqrt{-g} {R},
\end{equation}
is justified mainly by a criterion of simplicity rather than by fundamental principles. 
We have already briefly reviewed the evidence that a quantum theory of gravity will require a generalization of this action but indeed there has also been earlier phenomenological interest in this matter (see for example \cite{staro,buh,bar1,bar2}) and more recently, theories of gravity in which the Lagrangian includes additional terms with inverse powers of the scalar curvature have  also received considerable attention, since it was shown that they could account for the current accelerated expansion of the universe \cite{capo,capo2,carroll,odirev}. 

More generally, in a purely phenomenological approach, one might imagine replacing the scalar curvature in the Einstein--Hilbert action by  some function of it, $f(R)$ which could then be expanded in a power series (with positive as well as negative powers of the curvature scalar).
\begin{equation}
f(R)=\cdots +\frac{\alpha_2}{R^2}+\frac{\alpha_1}{R}-2\Lambda+R+\frac{R^2}{\beta_2}+\frac{R^3}{\beta_3}\cdots,
\label{eq:fr}
\end{equation}
where the $\alpha_i$ and $\beta_j$ coefficients have the appropriate dimensions.
In general, most of the research has been focused on the smallest deviations from the linear term in $R$, that is on Lagrangians with an extra $1/R$ or $R^2$ dependence. 

Unfortunately, all of the above theories (\ref{eq:fr}), usually called $f(R)$ theories of gravity, are riddled by problems. Apart from the increased complexity they exhibit, since they lead to fourth order field equations, they are also often burdened with some unwanted behaviour: {\em e.g.}~for the case of $1/R$ corrections (which  were proposed to explain the observed late time cosmological expansion) it is not clear whether simple models have the correct Newtonian limit \cite{dick,olmo,sot1,cem} and they do not seem to pass the solar system tests \cite{cem,chiba}. Even though it seems possible to construct a model that can avoid these issues \cite{noji}, this would require significant fine tuning of various parameters which does not seem appealing. What is more important though, is that, as shown in \cite{dolgov}, any model with a $1/R$ correction leads to unavoidable instabilities within matter in the weak gravity regime .

 Apart from the form of the action, another crucial assumption of General Relativity is that spacetime is fully described by the metric alone. Therefore the metric is considered to be the only fundamental field in the gravitational action. The necessity for this assumption was questioned very early, surprisingly by Einstein himself \cite{ffr}. The proposed alternative is to consider the metric and the affine connections as independent quantities, and vary the action with respect to both of them in order to derive the field equations. This of course implies that the connections are not chosen to be a priori the Levi--Civita connections of the metric. This approach is often called the Palatini formalism, even though Palatini was not the person who introduced it, but we shall use here the more precise terminology of the metric-affine formalism. 

\subsection{Motivations for the metric-affine formalism}
\label{sec:MAF}

This alternative formulation for gravity has been studied many times in the past (see for example \cite{pap,kun,tsamp,hehl,hehl2} and references therein\footnote{Note that the metric-affine approach has also been widely used in order to interpret gravity as a gauge theory (see for example \cite{rub} for a study on $f(R)$ actions and \cite{hehlrev2} for a thorough review).}). However, many of this earlier studies now seem slightly biased in their scope since, at that time, there was no motivation for questioning the validity of the Einstein field equations and the metric-affine formalism was merely viewed as an alternative way to derive them.  In this sense, any deviation away from standard GR was considered more as a drawback than as an opportunity, while, nowdays, modifications are instead welcome (if not actually pursued) for the reasons discussed above.

It is well known that the metric-affine formalism leads to the Einstein equations, when one starts from the Einstein--Hilbert action and assumes no dependence of the matter action on the connections \cite{wald}.  It has also been shown that, in vacuum, any $f(R)$ theory of gravity, treated within this framework, will lead to the Einstein field equations with an undetermined cosmological constant \cite{ferr,tap}. Recently, it was shown that adopting the metric-affine formalism together with an action that includes a term inversely proportional to the scalar curvature, such as the one in \cite{carroll}, can address the problem of the current accelerated expansion equally well as when using the purely metric formalism \cite{vollick}. Additionally, it was soon found out that $f(R)$ theories of gravity in the metric-affine formalism do not suffer from the problems discussed before for the metric formalism. 

More specifically, since in vacuum they straightforwardly reduce to standard General Relativity plus a cosmological constant they preserve interesting aspects of General Relativity, such as static black holes and gravitational waves. They are also free of the instabilities discovered in \cite{dolgov} for $f(R)$ gravity in the metric approach. Finally, even though there was initially
some debate concerning the Newtonian limit \cite{meng2,barraco}, it has been shown that they have the correct behavior \cite{sot1,gianluca}. Many other models \cite{meng,mw,mw2,meng3,gianluca2,sot2,sot3} have followed after the one presented in \cite{vollick}.

 It is interesting to note however that all the above mentioned models have an implicit assumption consisting in using a matter action which does not depend on the connections but only on the metric. This assumption is rather restrictive since in general the matter action can include covariant derivatives of the fields (at least if the minimal coupling principle, $\partial\to\nabla$, is naively applied). Of course, the recent rediscovery of the metric-affine (Palatini) formulation was mainly driven by the interest in finding cosmological scenarios able to explain the current observations. In these cases one deals mainly with matter in the form of a perfect fluid and hence with an action which is naturally independent of the connections. However the relevance of the metric-affine formulation as an alternative theory of gravity goes well beyond cosmological models and hence possible deviations from GR should be explored in other regimes as well.
Therefore we feel that it is urgent to pursue a deeper understanding of metric-affine gravity for more general gravitational and matter Lagrangians.

Metric-affine gravity in the presence of matter whose action also depends on the connections has been studied in the past. In \cite{hehl,hehl2} the Einstein--Hilbert action was considered allowing also torsion. In \cite{mang} $f(R)$ theories were studied in the absence of torsion. What we would like to study here is the full version of the theory, {\em i.e.}~$f(R)$ theories of gravity in the metric-affine formalism and in the presence of matter, allowing inclusion also of torsion, by dropping the assumption that the connection is necessarily symmetric.

\subsection{Plan of the paper}
\label{sec:plan}
We start in section \ref{form} by stating the basic definitions and conventions which we shall be using and then discuss the form used for the gravitational action and the physical motivation for this choice. We then proceed, in section \ref{sec:MAvac}, by reviewing the derivation of the gravitational field equations in vacuum. The reason for this is that we want to highlight some points that have been overlooked, or not given enough attention in the past literature. 
In section \ref{matter} we develop the formalism for metric-affine theories in the presence of matter. We allow the matter Lagrangian to depend not only on the metric but also on the connections, which as we argue is the most natural choice. This is done with and without torsion; we shall see that in the presence of torsion 
a thorough and rigorous treatment is necessary, since many subtleties appear. In section \ref{sec:MaAct} we study the physical interpretation of the dependence of the matter action on the connections and consider different matter configurations. We show that this dependence is essential for torsion and relates its presence to matter fields that describe particles with spin (apart from the case of electromagnetic/gauge fields that we discuss in detail). In addition, we shall see that whenever the matter fields are such that they do not depend on the connection, torsion automatically vanishes.

Another problem that our analysis  allows us to address is the debate about whether metric-affine $f(R)$ theories of gravity  are in conflict with the standard model of particle physics \cite{flan,voll1,flan2,voll2}. It has been proved that this problem arises when one assumes that the matter action does not depend on the connection \cite{flan}. However, as shown in \cite{voll2} this is no longer the case in the full version of the theory which we are considering here. We discuss that issue here and examine its implications for the cosmological models presented in the literature (see {\em e.g.}~\cite{vollick,meng,mw,mw2,meng3,sot2}). In section \ref{conc} we present conclusions.

\section{Metric-affine formalism}
\label{form}

We would like to start by presenting the formalism of metric-affine gravity which we shall use throughout this paper. This will allow  us to construct the gravitational  action, the variation of which will lead to the field equations. We consider a 4-dimensional manifold with a connection, $\Gamma_{\phantom{a}\mu\nu}^\lambda$, and a symmetric metric $g_{\mu\nu}(=g_{\nu\mu})$. In such a manifold the metric allows us to measure distances and defines the causal structure  while the connection is related to parallel transport and therefore defines the covariant derivative. The definition is the following:
\be
\nabla_\mu A^\nu_{\phantom{a}\sigma}=\partial_\mu A^\nu_{\phantom{a}\sigma}+\Gamma^\nu_{\phantom{a}\alpha\mu} A^\alpha_{\phantom{a}\sigma}-\Gamma^\alpha_{\phantom{a}\mu\sigma} A^\nu_{\phantom{a}\alpha}\, ,
\ee
(we give it here even though it may be considered trivial, since several different conventions exist in the literature. Additionally one has to be careful about the position of the indices when the connection  is not symmetric.) Using the connection one can construct the Riemann tensor:
\be
\label{riemann}
R^\mu_{\phantom{a}\nu\sigma\lambda}=-\partial_\lambda\Gamma^\mu_{\phantom{a}\nu\sigma}+\partial_\sigma\Gamma^\mu_{\phantom{a}\nu\lambda}+\Gamma^\mu_{\phantom{a}\alpha\sigma}\Gamma^\alpha_{\phantom{a}\nu\lambda}-\Gamma^\mu_{\phantom{a}\alpha\lambda}\Gamma^\alpha_{\phantom{a}\nu\sigma}\, .
\ee
which has no dependence on the metric. Notice that the Riemann tensor here has only one obvious symmetry; it is antisymmetric in the last two indices. The rest of the standard symmetries are not present for an arbitrary connection \cite{schro}.

In the case of a purely metric theory, one makes the assumption that the affine connections $\Gamma_{\phantom{a}\mu\nu}^\lambda$ are the Levi--Civita connections, {\em i.e.}~$\Gamma_{\phantom{a}\mu\nu}^\lambda=\{_{\phantom{a}\mu\nu}^\lambda\}$. 
This assumption is actually dual, since it requires both the metric to be covariantly conserved ($g_{\mu\nu;\lambda}$=0) and the connection to be symmetric ($\Gamma_{\phantom{a}\mu\nu}^\lambda=\Gamma_{\phantom{a}\nu\mu}^\lambda$). With a semicolon we denote covariant differentiation with respect to the connections $\Gamma_{\phantom{a}\mu\nu}^\lambda$. Given these assumptions the Riemann tensor will turn out to be antisymmetric  also with respect to the first two indices as well as symmetric in an exchange of the first and the second pairs. Therefore, one can construct only one second rank tensor from straightforward contraction, {\em i.e.}~without using the metric. This is the well known  Ricci tensor. A full contraction with the metric will then lead to the Ricci scalar in the usual way. In order to construct an action whose variation leads to second order field equations we need to find a generally covariant scalar that depends on the metric and its first derivatives only. Unfortunately, as is well known, no such scalar exists. The simplest generally covariant scalar that one can construct is the Ricci scalar which depends also on the second derivatives of the metric. This was the motivation of Hilbert for defining the gravitational action of a purely metric theory as:
\be
\label{actioneh}
S_{EH}=\frac{1}{2\kappa}\int d^4 x \sqrt{-g} \stackrel{g}{R},
\ee
where $g$ denotes the determinant of the metric $g_{\mu\nu}$ and $\stackrel{g}{R}$ denotes the Ricci scalar. The $g$ placed over $R$ is used as a reminder for the fact that, in this case, the Ricci scalar is a function of the metric. Even though $\stackrel{g}{R}$ depends on the second derivatives of the metric this action leads to second order field equations since the terms including the second derivatives can be collected in a total divergence, {\em i.e.}~a surface term. We shall come back to this later.

Things are not equally straightforward in the Palatini formalism as we shall see. We do not assume here any relation between the metric and the connections. This means  that the metric in not necessarily covariantly conserved. The failure of the connection to preserve the metric is usually measured by the non-metricity tensor:
\be
\label{nonmet}
Q_{\mu\nu\lambda}\equiv-\nabla_\mu g_{\nu\lambda}.
\ee
The trace of the non-metricity tensor with respect to its last two (symmetric) indices is called the Weyl vector:
\be
\label{weyl}
Q_\mu\equiv \frac{1}{4}Q_{\mu\nu}^{\phantom{a}\phantom{b}\nu}.
\ee
At the same time the connection is not necessarily symmetric. The antisymmetric part of the connection is often called the Cartan torsion tensor:
\be
\label{cartan}
S_{\mu\nu}^{\phantom{ab}\lambda}\equiv \Gamma^{\lambda}_{\phantom{a}[\mu\nu]}.
\ee

One has to be careful when  deriving the Ricci tensor in this case, since only some of the standard symmetry properties of the Riemann tensor 
hold here. A straightforward contraction leads,  in fact, to two Ricci tensors \cite{schro}:
\be
R_{\mu\nu}\equiv R^\sigma_{\phantom{a}\mu\sigma\nu}, \qquad R'_{\mu\nu}\equiv R^\sigma_{\phantom{a}\sigma\mu\nu}.
\ee
The first one is usual Ricci tensor given by
\be
\label{ricci}
R_{\mu\nu}=R^\lambda_{\phantom{a}\mu\lambda\nu}=\partial_\lambda \Gamma^\lambda_{\phantom{a}\mu\nu}-\partial_\nu \Gamma^\lambda_{\phantom{a}\mu\lambda}+\Gamma^\lambda_{\phantom{a}\sigma\lambda}\Gamma^\sigma_{\phantom{a}\mu\nu}-\Gamma^\lambda_{\phantom{a}\sigma\nu}\Gamma^{\sigma}_{\phantom{a}\mu\lambda}.
\ee 
The second is given by the following equation
\be
R'_{\mu\nu}=-\partial_\nu \Gamma^\alpha_{\phantom{a}\alpha\mu}+\partial_\mu \Gamma^\alpha_{\phantom{a}\alpha\nu}.
\ee
For a symmetric connection this tensor is equal to the antisymmetric part of $R_{\mu\nu}$. Fully contracting both tensors with the metric to get a scalar gives, for $R_{\mu\nu}$
\be
\label{rscal}
R=g^{\mu\nu}R_{\mu\nu}
\ee
which is the Ricci scalar, and for $R'_{\mu\nu}$
\be
R'=g^{\mu\nu}R'_{\mu\nu}=0,
\ee
since the metric is symmetric and $R'_{\mu\nu}$ antisymmetric. Therefore the Ricci scalar is uniquely defined from eq.~(\ref{rscal}) \footnote{We considered second rank tensors that one gets from a contraction of the Riemann tensor without using the metric, {\em i.e.}~tensors independent of the metric. There is a third second rank tensor which can be built from the Riemann tensor \cite{tsamp}: $R''_{\mu\nu}\equiv R^{\phantom{a}\sigma}_{\mu\phantom{a}\sigma\nu}=g^{\sigma\alpha}g_{\mu\beta}R^{\beta}_{\phantom{a}\alpha\sigma\nu}$. This tensor, however, depends on the metric. A further contraction with the metric will give $R''=g^{\mu\nu}R''_{\mu\nu}=-R$, and so even if we consider this tensor, the Ricci scalar is uniquely defined.}.

Having established this subtle point, we now proceed in constructing the gravitational action. In the process of deriving the Einstein--Hilbert action, eq.~(\ref{actioneh}), we considered only $\s{g}{R}$ motivated initially by the fact that we want the resulting field equations to be second order differential equations. This demand comes from the fact that all other theories besides gravity are described by such field equations. Using the same demand, we can build the action for metric-affine gravity. We need a generally covariant scalar that depends only on our fundamental fields, the metric and the connections, and their first derivatives at most. Therefore the obvious choice is the Ricci scalar $R$. Actually, there is no other generally covariant scalar with these properties. 

Of course, one could try to use $\s{g}{R}$, {\em i.e.}~the scalar curvature related to the metric alone. Another option can arise if the connections are of such a form that one can define a second metric, $h_{\mu\nu}$, that is covariantly conserved, {\em i.e.}~the metric of which the $\Gamma$'s are the Levi--Civita connections (note that this is not necessarily true for a general connection~\cite{Gm}, so it would lead to a less general theory). Then we could use this metric to contract the Riemann tensor and derive the Ricci scalar, $\stackrel{h}{R}$, which is actually the scalar curvature of the metric $h_{\mu\nu}$ . Going even further we could even use one of the two metrics, $g_{\mu\nu}$ or $h_{\mu\nu}$, to go from the Riemann tensor to the Ricci tensor and the other to derive the Ricci scalar from the Ricci tensor. The question that arises is whether not using these other scalar quantities in the action constitutes a further assumption, which is not needed in the purely metric formulation. 

 From the mathematical point of view, we could use any of the Ricci scalars defined above. However we think that for any possible choice other than (\ref{rscal}) there are good physical reasons for discarding it.
In fact when constructing a metric-affine theory, one assumes that the spacetime is fully described by two independent geometrical objects, the metric and the connection. The metric defines the chronological structure, the connection defines the affine structure of the manifold. This manifold is not chosen to be Riemannian (at least initially). One can always mathematically consider two Riemannian manifolds, one described by the metric $g_{\mu\nu}$ and the other by the metric $h_{\mu\nu}$ (if it exists), but these manifolds are not relevant for the spacetime in which the theory acts. Therefore, quantities related to them, like their scalar curvatures, should not be used in the action of a theory living on the non-Riemannian manifold under consideration. Additionally, using quantities derived by contracting once with one metric and once with the other, should also be avoided. There is only one metric that determines how distances are measured in our spacetime and this is $g_{\mu\nu}$. This is the metric that is used to evaluate inner products and therefore is the one that should be used to raise or lower indices and perform contractions.

Given the above discussion, we conclude that the most 
natural choice in metric-affine theories is to use $R$ as the Lagrangian of our action and write
\be
\label{action}
S=\frac{1}{2\kappa}\int d^4 x \sqrt{-g} R.
\ee
Notice the following however. Our demand for second order field equations allows us to use any function $f(R)$ as a Lagrangian, since any $f(R)$ will be independent of derivatives higher than first order. Therefore, choosing an action linear in $R$, like (\ref{action}) must be considered as a simplifying choice in metric-affine gravity, unlike in purely metric theories where an action linear in $R$ is the only one that leads to second order equations \footnote{We are confining ourselves to Lagrangians that are functions of the Ricci scalar only. In a more general setting one should mention that Gauss-Bonnet type Lagrangians lead to second order field equations as well.}. In metric-affine theories of gravity one can say that $f(R)$ actions are as ``natural'' as the Einstein--Hilbert one.

\section{Metric-affine gravity in vacuum}
\label{sec:MAvac}

Before studying the full version of metric-affine gravity we would like to  consider in this section the gravitational action in the absence of matter, under the simplifying assumption that the connection is symmetric ($\Gamma_{\phantom{a}\mu\nu}^\lambda=\Gamma_{\phantom{a}\nu\mu}^\lambda$). As mentioned in the introduction, this simplified version of the theory has been studied before. However, we are repeating this study here since, besides being an intermediate step towards the full theory, it will also give us a chance to clarify many subtle points in the past literature. We shall start with an action which is linear in the scalar curvature and subsequently generalize our results for the non-linear case.

\subsection{Linear action}
\label{sec:linear}

Let us now see how one derives the field equations from the actions (\ref{actioneh}) and (\ref{action}) {\em i.e.}~in the metric and metric-affine frameworks respectively. We shall not discuss these cases in detail however, since they are standard text book results (see {\em e.g.}~\cite{wald}). 

The variation of the action (\ref{actioneh}) with respect to the metric gives
\be
\label{varm}
0=\frac{1}{2\kappa}\left[\int_U d^4 x \sqrt{-g} G_{\mu\nu}\delta g^{\mu\nu}-2\int_{\delta U} d^3 x \sqrt{
|h|}\, \delta K\right],
\ee
where $U$ denotes the volume, $\delta U$ denotes the boundary of $U$, and $K$ is as usual the trace of the extrinsic curvature of that boundary. $G_{\mu\nu}\equiv \stackrel{g}{R_{\mu\nu}}-\frac{1}{2}g_{\mu\nu}\stackrel{g}{R}$ is the Einstein tensor. The second term in eq.~(\ref{varm}) is a surface term. Assuming that $g_{\mu\nu}$ is fixed on the boundary does not imply, however, that this term goes to zero. That would require also the first derivatives of the metric to be fixed on the boundary. Therefore in order to properly derive the Einstein equations one has to redefine the gravitational action as
\be
\label{actionehm}
S'_{EH}=S_{EH}+\frac{1}{\kappa}\int_{\delta U}d^3 x \sqrt{|h|}\, K.
\ee
Using this action one has cancellation of the surface terms and hence a clean derivation of the Einstein field equations (for a detailed discussion of the role and nature of the surface term in (\ref{actionehm}) see {\em e.g.}~\cite{paddy}).

Passing to the metric-affine framework, the variation of the action (\ref{action}) should now be done with respect to both the metric and the connections (or the covariant derivatives) separately. The easiest way to do this is to follow \cite{wald} and express the $\Gamma$'s, as a sum of the  Levi--Civita connections of the metric, $g_{\mu\nu}$,  and a tensor field $C_{\phantom{a}\mu\nu}^\lambda$. Variation with respect to the $\Gamma$'s (or the covariant derivative) will then be equivalent to the variation of $C_{\phantom{a}\mu\nu}^\lambda$. On the boundary, $g_{\mu\nu}$ and $C_{\phantom{a}\mu\nu}^\lambda$ will be fixed and we get the following:
\bea
\label{varp}
0&=&-2\frac{1}{2\kappa}\int d^4 x\sqrt{-g} g^{\mu\nu}\tilde{\nabla}_{[\mu}\delta C_{\phantom{a}\lambda]\nu}^\lambda+{}\nn\\
& &{}+\frac{1}{2\kappa}\int d^4 x\sqrt{-g} \left(C^{\nu\sigma}_{\phantom{a}\phantom{a}\sigma}\delta^\mu_{\phantom{a}\lambda}+C^\sigma_{\phantom{a}\sigma\lambda}g^{\mu\nu}-2C^{\nu\phantom{a}\mu}_{\phantom{a}\lambda}\right)\delta C^\lambda_{\phantom{a}\mu\nu}+\nn\\
& &{}+\frac{1}{2\kappa}\int d^4 x\sqrt{-g} \left(R_{\mu\nu}-\frac{1}{2}Rg_{\mu\nu}\right)\delta g^{\mu\nu},
\eea
where $\tilde{\nabla}$ denotes the covariant derivative related to the Levi-Civita connection. We see immediately that the first term in eq.~(\ref{varp}) is again a surface term. This time, however, it is exactly zero, since now $\delta C_{\phantom{a}\mu\nu}^\lambda=0$ on the boundary as $C_{\phantom{a}\mu\nu}^\lambda$ is fixed there\footnote{The fact that the surface terms vanish on the boundary in metric-affine gravity is one of its advantages with respect to purely metric theories. As we saw earlier for the Einstein--Hilbert action one gets a non-vanishing surface term and a modification of the initial action is required to allow the derivation of the field equations. Note however that, even though this is a standard procedure for the Einstein--Hilbert action, it is not guaranteed to work for more general actions.}. 

 Coming back to (\ref{varp}), we see now that asking for the second term to vanish corresponds to the condition  
\be
C_{\phantom{a}\mu\nu}^\lambda=0,
\ee
or
\be
\label{f1}
\Gamma_{\phantom{a}\mu\nu}^\lambda=\{_{\phantom{a}\mu\nu}^\lambda\},
\ee
{\em i.e.}~the $\Gamma$'s  have to be the Levi--Civita connections of the metric. 
So, in the end, the last term leads to the standard Einstein equations given that now, due to eq.~(\ref{f1}), $R_{\mu\nu}=\s{g}{R}_{\mu\nu}$. Note that the above results remain unchanged if a cosmological constant is added to the action as the resulting equations will be just the standard Einstein equations with a non-vanishing cosmological constant.

It is worth making the following comment: in the metric variation one gets a non zero surface term. In order to get the Einstein equations one has, therefore, to modify the action. This is not true for the metric affine variational principle. The part of the action that contributes the surface term in the metric approach, is now split into two parts. One of these leads to eq.~(\ref{f1}), and the other turns out to give again a surface term, which however is zero. The covariant forms of actions (\ref{actioneh}) and (\ref{action}) coincide. However this is not true for the actions (\ref{actionehm}) and (\ref{action}), so {\it we cannot say that the Einstein--Hilbert action leads to the Einstein equation with both variational principles} as it is usually said. {\it The Einstein equations are derived with both variational principles, but starting from actions that differ by a surface term}.

Another way to see this is the following. Choosing the purely metric variational principle means that we are assuming that the fundamental field related to gravity is $g_{\mu\nu}$. However, under this assumption, the Einstein--Hilbert actio, has a rather unusual attribute; it depends not only on the first derivative of this field, but also on the second derivatives, unlike in most theories. Therefore, it should not straightforwardly lead to second order field equations. Subtracting the extrinsic curvature term from the action as in (\ref{actionehm}), one eliminates the dependence on the second order derivatives, and therefore obtains second order field equations. On the other hand, when we choose a metric-affine variational principle, we actually assume that there are two fundamental fields in our action: the metric and the connections. Under this assumption, the Einstein--Hilbert action does not depend on the second derivatives of the metric, but only on the first derivatives of the metric and the first derivatives of the connections. Therefore, it will indeed straightforwardly lead to second order field equations for both fields, without the need for a surface term. From this point of view, the metric-affine formulation of gravity has a serious advantage: the gravitational action depends only on the first derivatives of fundamental fields, in close analogy with all other commonly used theories (such as Electrodynamics, for example).

\subsection{Non-Linear Action}
\label{nonlinear}

Let us now drop the assumption that the gravitational action has to be linear in the scalar curvature. The action will then be a general function of $R$:
\be
\label{nlaction}
S=\frac{1}{2\kappa}\int d^4x \sqrt{-g} f(R).
\ee
The variation of the action with respect to the metric will give
\be
\label{e1}
f'(R) R_{\mu\nu}-\frac{1}{2}f(R)g_{\mu\nu}=0,
\ee
and the variation with respect to the connections will give
\be
\label{e21}
\nabla_\lambda\left(f'(R)\sqrt{-g}g^{\mu\nu}\right)-\nabla_\sigma \left(f'(R)\sqrt{-g} 
g^{\sigma(\mu}\right){\delta^{\nu)}}_{\lambda}=0.
\ee
By taking the trace of eq.~(\ref{e21}) we get
\be
\label{e2}
\nabla_\lambda\left(f'(R)\sqrt{-g}g^{\mu\nu}\right)=0.
\ee
This equation tells us that we can in general define a metric,
\be
\label{confm}
h_{\mu\nu}=f'(R)g_{\mu\nu},
\ee
which will be covariantly conserved. Therefore, the affine connections will be the Levi--Civita connections of $h_{\mu\nu}$. 

If we take the trace of eq.~(\ref{e1}) we get
\be
\label{scalar}
f'(R) R-2 f(R)=0.
\ee
This is an algebraic equation for $R$ once $f(R)$ has been specified. In general we expect this equation to have a number of solutions,
\be
\label{sol}
R=c_i,\quad i=1,2,\dots
\ee
were the $c_i$ are constants.
There is also a possibility that eq.~(\ref{scalar}) will have no real solutions or will be satisfied for any $R$ (which happens for $f(R)=a R^2$, where $a$ is an arbitrary constant). In the first case there are no consistent field equations \cite{ferr}. In the second case, the field equations will be consistent in vacuum, but notice the following: if one tries to add matter to such a model, then the right hand side of eq.~(\ref{scalar}) will turn out to be the trace of the stress-energy tensor (defined in the usual way). Then the vanishing of the left hand side would imply that the trace of the stress-energy tensor has to be zero, which of course is not true for all matter fields and therefore leads to an inconsistency. Since these cases constitute exceptions that mainly seem uninteresting or are burdened with serious difficulties when matter is also considered, we shall not study them here. They were studied to some extent in \cite{ferr}. 

Let us, therefore, return to the case where eq.~(\ref{scalar}) has the solutions given in eq.~(\ref{sol}). In this case, since $R$ is a constant, then $f'(R)$ is also a constant and eq.~(\ref{e2}) reads
\be
\label{e3}
\nabla_\lambda\left(\sqrt{-g}g^{\mu\nu}\right)=0.
\ee
This is the metricity condition for the affine connections, $\Gamma^\lambda_{\phantom{a}\mu\nu}$. Thus, the affine connections now become the Levi--Civita connections of the metric, $g_{\mu\nu}$. Eq.~(\ref{e1}) then reads
\be
\label{e4}
R_{\mu\nu}-\frac{1}{4}c_i g_{\mu\nu}=0,
\ee
with 
\be
\label{e5}
\Gamma_{\phantom{a}\mu\nu}^\lambda=\{_{\phantom{a}\mu\nu}^\lambda\},
\ee
which is exactly the Einstein field equation with a cosmological constant.

Therefore, in the end we see that a general $f(R)$ theory of gravity in vacuum, studied within the framework of metric-affine variation, will lead to the Einstein equation with a cosmological constant.  This is not the case if one uses the metric variational principle as, in this case, one ends up with fourth order field equations, {\em i.e.}~with a significant departure from the standard Einstein equations (see for example \cite{buh}).
Another important feature that deserves to be commented upon is the following: Contrary to the spirit of General Relativity where the cosmological constant has a unique value, here the cosmological constant is also allowed to have different values, $c_i$, corresponding to different solutions of eq.~(\ref{scalar}).  So, in a sense, the action (\ref{nlaction}) is equivalent to a whole set of Einstein--Hilbert actions \cite{mang} (or, more precisely, actions of the form (\ref{actionehm}) plus a cosmological constant).

\section{Metric-affine gravity with matter}
\label{matter}

In this section we consider an action that includes, besides gravity, also matter fields. This time we shall not consider separately the case of a linear gravitational action as we did before, however all the results presented below apply also to linear Lagrangians, such as $f(R)=R$ or $f(R)=R-2 \Lambda$, where $\Lambda$ is a cosmological constant.

\subsection{The variation of the action}
\label{variation}

One of the key assumptions of General Relativity is that the metric is a symmetric, non-singular, tensor. Since the connections are also assumed to be the the Levi--Civita connections of this metric, they are automatically symmetric as well. However, for a metric-affine theory of gravity this is not necessarily true. Since the metric and the connections are independent, the metric can be symmetric without the connections being symmetric as well.  In the previous section we had explicitly assumed symmetric connections for simplicity (for further discussion about $f(R)$ metric-affine gravity without torsion see also \cite{mang,ferr}). From here on, we are going to drop this assumption, so that we can develop a more general formalism for metric-affine theories of gravity. Given that the anti-symmetric part of the connection is related to torsion, our choice implies the possibility of emergence of the latter in some situations. Metric-affine gravity with a non symmetric connection has also been considered in \cite{hehl} for linear actions only and so the formalism developed in this section is also a generalization of the result presented there.

The full action has the form
\be
\label{fulllang}
S=S_G+S_M,
\ee
where
\be
S_G=\frac{1}{2\kappa}\int d^4x \sqrt{-g} f(R),
\ee
 and $S_M$ is the matter action whose detailed form can be left unspecified for the moment. The least action principle gives
\be
\label{var}
0=\delta S=\delta S_G+\delta S_M,
\ee
and the variation of the gravitational part gives
\bea
\label{varg}
\delta S_G&=&\frac{1}{2\kappa}\int d^4 x \,\delta\left(\sqrt{-g}f(R)\right)=\nn\\
&=&\frac{1}{2\kappa}\int d^4 x \left(f(R)\delta\sqrt{-g}+\sqrt{-g}f'(R)\delta R\right)\nn\\
&=&\frac{1}{2\kappa}\int d^4 x \left(f(R)\delta\sqrt{-g}+\sqrt{-g}f'(R)\delta \left(g^{\mu\nu}R_{\mu\nu}\right)\right)\nn\\
&=&\frac{1}{2\kappa}\int d^4 x \sqrt{-g}\left(f'(R) R_{(\mu\nu)}-\frac{1}{2}f(R)g_{\mu\nu}\right)\delta g^{\mu\nu}+\nn\\& &+\frac{1}{2\kappa}\int d^4 x \sqrt{-g}f'(R)g^{\mu\nu}\delta R_{\mu\nu},
\eea
where we have used the symmetry of the metric ($\delta g^{\mu\nu}R_{\mu\nu}=\delta g^{\mu\nu}R_{(\mu\nu)}$). 

To complete this variation we need to evaluate the quantity $\delta R_{(\mu\nu)}$. $R_{\mu\nu}$ depends only on the connections and so we can already see that the second term of the last line of eq.~(\ref{varg}) will be the one related to the variation with respect to $\Gamma^\lambda_{\phantom{a}\mu\nu}$.
Using the definition of the Ricci tensor, eq.~(\ref{ricci}), one can show after some mathematical manipulations that
\bea
\label{varR}
\delta R_{\mu\nu}&=&\nabla_\lambda \delta \Gamma^\lambda_{\phantom{a}\mu\nu}-\nabla_{\nu}\delta \Gamma^\lambda_{\phantom{a}\mu\lambda}+2\Gamma^\sigma_{\phantom{a}[\nu\lambda]}\delta\Gamma^\lambda_{\phantom{a}\mu\sigma}.
\eea
Using eq.~(\ref{varR}) the variation of the gravitational part of the action takes the form
\bea
\label{varg2}
\delta S_G&=&\frac{1}{2\kappa}\int d^4 x \sqrt{-g}\left(f'(R) R_{(\mu\nu)}-\frac{1}{2}f(R)g_{\mu\nu}\right)\delta g^{\mu\nu}+{}\nn\\& &+\frac{1}{2\kappa}\int d^4 x \sqrt{-g}f'(R)g^{\mu\nu}\left(\nabla_\lambda \delta \Gamma^\lambda_{\phantom{a}\mu\nu}-\nabla_{\nu}\delta \Gamma^\lambda_{\phantom{a}\mu\lambda}\right)+\nn\\& &+\frac{1}{2\kappa}\int d^4 x \,2\sqrt{-g}f'(R)g^{\mu\sigma}\Gamma^\nu_{\phantom{a}[\sigma\lambda]}\delta\Gamma^\lambda_{\phantom{a}\mu\nu}.
\eea
Integrating the terms in the second line by parts we get
\bea
\label{varg3}
\delta S_G&=&\frac{1}{2\kappa}\int d^4 x \sqrt{-g}\left(f'(R) R_{(\mu\nu)}-\frac{1}{2}f(R)g_{\mu\nu}\right)\delta g^{\mu\nu}+{}\\& &+\frac{1}{2\kappa}\int d^4 x \bigg[-\nabla_\lambda\left(\sqrt{-g}f'(R)g^{\mu\nu}\right)+\nabla_\sigma\left(\sqrt{-g}f'(R)g^{\mu\sigma}\right)\delta^\nu_\lambda\nn\\& &+ 2\sqrt{-g}f'(R)\left(g^{\mu\nu}\Gamma^\sigma_{\phantom{a}[\lambda\sigma]}-g^{\mu\rho}\Gamma^\sigma_{\phantom{a}[\rho\sigma]}\delta^\nu_\lambda+g^{\mu\sigma}\Gamma^\nu_{\phantom{a}[\sigma\lambda]}\right)\bigg]\delta\Gamma^\lambda_{\phantom{a}\mu\nu}+\textrm{ST},\nn
\eea
where $\textrm{ST}$ stands for ``Surface Terms''. These terms are total divergences linear in $\delta \Gamma^\lambda_{\phantom{a}\mu\nu}$. Being total divergences, we can turn their integral over the volume into an integral over the boundary surface. Since $\delta \Gamma^\lambda_{\phantom{a}\mu\nu}=0$ on the boundary, they will vanish. [Note that the first two terms in the last line of eq.~(\ref{varg3}) came from the integration by parts of the second line of (\ref{varg2}). This is because differentiation by parts and integration of covariant derivatives becomes non-trivial in the presence of a non-symmetric connection (for more information on this, see \cite{schro}, chapter 2 and  p.~109).] This concludes the variation of the gravitational part of the action.

We now have to consider the variation of the matter action. Before doing this we need to clarify its dependence on the gravitational field.  In a large portion of the literature, mostly related to cosmological applications, $S_M$ is taken to be independent of the connections (see, for example, \cite{sot1,vollick,meng3,meng,sot2,mw,mw2}). However, in general $S_M$ can depend on both the metric and on the covariant derivative. This practically means that, in the theory studied in the papers mentioned previously, the covariant derivative used in the matter action is chosen to be {\it a priori} the metric compatible one. Therefore, in such theories the affine connection used to construct the action does not carry the usual geometrical interpretation. Neither does it define parallel transport, nor is it at all relevant to the geometrical stucture of spacetime, which is actually assumed {\it a priory} to be described by the metric alone, even though this might not be obvious. Hence we can conclude that this connection is demoted to being a sort of an auxiliary field,  which, furthermore, is not coupled to matter. This is equivalent to saying that such a theory is intrinsically a metric theory of gravity.

To clarify the above discussion, let us comment that one can give to this connection a similar physical interpretation to the one given to the scalar field in scalar-tensor theory. The metric is the only part of the gravitational field that interacts with matter. The scalar field just intervenes in the generation of the spacetime curvature induced by the matter fields and associated with the metric without of course affecting the geometry. The $\Gamma$s, when they are not coupled to the matter, work in exactly the same manner, and simply participate in the way matter tells spacetime how to curve, without actually carrying any characteristics of the curvature themselves. It has been shown that, for such matter coupling, the discussed theory can indeed be cast into the form of a Brans--Dicke theory \cite{olmo}. We will call $f(R)$ theories in which the matter action is chosen to be independent of the connection, $f(R)$ theories of gravity in the Palatini formalism, to make the distinction from metric-affine $f(R)$ gravity.

In the present paper we started by the assumption that the connections defines parallel transport and consequently the covariant derivative in the gravitational sector. Forcing the matter action to be independent of the connections would, as we said, contradict this assumption.
Therefore we conclude that in a truly metric-affine theory of gravity the matter action should be allowed to depend on the connections in general. Hence, $S_M=S_M(g_{\mu\nu},\Gamma^\lambda_{\phantom{a}\mu\nu},\psi)$, where $\psi$ represents the matter fields. We then have 
\be
\delta S_M=\int d^4 x \frac{\delta S_M}{\delta g^{\mu\nu}}\delta g^{\mu\nu}+\int d^4 x\frac{\delta S_M}{\delta \Gamma^\lambda_{\phantom{a}\mu\nu}}\delta \Gamma^\lambda_{\phantom{a}\mu\nu}.
\ee
We can define the stress-energy tensor in the usual way
\be
\label{set}
T_{\mu\nu}\equiv-\frac{2}{\sqrt{-g}}\frac{\delta {S}_M}{\delta g^{\mu\nu}}.
\ee
We also define a new tensor, which we shall call (following the nomenclature of \cite{hehl}) the ``hypermomentum'', as
\be
\label{defD}
\Delta_{\lambda}^{\phantom{a}\mu\nu}\equiv-\frac{2}{\sqrt{-g}}\frac{\delta {S}_M}{\delta \Gamma^\lambda_{\phantom{a}\mu\nu}},
\ee
{\em i.e.}~the variation of the matter action with respect to the connections. Therefore, the variation of the matter action will be
\be
\label{varmat}
\delta S_M=-\frac{1}{2}\int d^4 x\sqrt{-g}\left[T_{\mu\nu}\delta g^{\mu\nu}+\Delta_{\lambda}^{\phantom{a}\mu\nu}\delta \Gamma^\lambda_{\phantom{a}\mu\nu}\right].
\ee

Note that the vanishing of $\Delta_{\lambda}^{\phantom{a}\mu\nu}$ would imply independence of the matter action from the connections. As we discussed this would be unphysical in the context of metric-affine gravity if it happend for any field (the theory would drop to $f(R)$ gravity in the Palatini formalism). There are, however, specific fields that have this attribute; the most common example is the scalar field. There will be therefore cases, as we will see later on, where metric-affine $f(R)$ gravity and $f(R)$ gravity in the Palatini formalism will give equivalent physical predictions, without of course being equivalent theories.  For instance, if we consider a massive vector field or a Dirac field, the matter action is no longer independent of the connection and  $\Delta_{\lambda}^{\phantom{a}\mu\nu}$ does not vanish. 

\subsection{The field equations}
\label{fieldeq}
 We are now ready to derive the field equations using the variation of the gravitational and matter actions. This can be achieved simply by summing the variations (\ref{varg3}) and (\ref{varmat}) and applying the least action principle. We obtain
\be
\label{field1}
f'(R) R_{(\mu\nu)}-\frac{1}{2}f(R)g_{\mu\nu}=\kappa T_{\mu\nu},
\ee
and
\bea
\label{field2}
\frac{1}{\sqrt{-g}}\bigg[&-&\nabla_\lambda\left(\sqrt{-g}f'(R)g^{\mu\nu}\right)+\nabla_\sigma\left(\sqrt{-g}f'(R)g^{\mu\sigma}\right){\delta^\nu}_\lambda\bigg]+{}\nn\\ & &+2f'(R)\left(g^{\mu\nu}\Gamma^\sigma_{\phantom{a}[\lambda\sigma]}-g^{\mu\rho}\Gamma^\sigma_{\phantom{a}[\rho\sigma]}{\delta^\nu}_\lambda+g^{\mu\sigma}\Gamma^\nu_{\phantom{a}[\sigma\lambda]}\right)=\kappa\Delta_{\lambda}^{\phantom{a}\mu\nu}.
\eea
We can also use the Cartan torsion tensor, eq.~(\ref{cartan}),
to re-express eq.~(\ref{field2}) and highlight the presence of torsion:
\bea
\label{field22}
\frac{1}{\sqrt{-g}}\bigg[&-&\nabla_\lambda\left(\sqrt{-g}f'(R)g^{\mu\nu}\right)+\nabla_\sigma\left(\sqrt{-g}f'(R)g^{\mu\sigma}\right){\delta^\nu}_\lambda\bigg]+{}\nn\\ & &+2f'(R)\left(g^{\mu\nu}S^{\phantom{ab}\sigma}_{\lambda\sigma}-g^{\mu\rho}S^{\phantom{ab}\sigma}_{\rho\sigma}{\delta^\nu}_\lambda+g^{\mu\sigma}S^{\phantom{ab}\nu}_{\sigma\lambda}\right)=\kappa\Delta_{\lambda}^{\phantom{a}\mu\nu}.
\eea
A careful look at the above equation reveals that if we take the trace on $\lambda$ and $\mu$ we get
\be
\label{contr}
0=\kappa \Delta_{\mu}^{\phantom{a}\mu\nu},
\ee
since the left hand side is traceless. One can interpret this as a constraint on the form of $\Delta_{\lambda}^{\phantom{a}\mu\nu}$, meaning that the matter action has to be chosen in such a way that its variation with respect to the connections leads to a traceless tensor. However, it is easy to understand that this is not too appealing as it restricts the forms of matter that our theory can describe. On the other hand we cannot assume that any form of matter will have this attribute. Therefore the field equations we have derived seem to be inconsistent. This problem is not new; it has been pointed out for the simple case of the Einstein--Hilbert action long ago \cite{hehl,schro,sand}. Its roots can be traced in the form of the action itself and in the fact that in metric-affine gravity $\Gamma^{\lambda}_{\phantom{a}\mu\nu}$ has no {\em a priori} dependence on the metric.

Let us consider the projective transformation
\be
\label{proj}
\Gamma^{\lambda}_{\phantom{a}\mu\nu}\rightarrow \Gamma^{\lambda}_{\phantom{a}\mu\nu}+{\delta^\lambda}_\mu\xi_\nu,
\ee
where $\xi_\nu$ is an arbitrary covariant vector field. One can easily show that the Ricci tensor will  correspondingly transform like
\be
\label{projRicci}
R_{\mu\nu}\rightarrow R_{\mu\nu}-2\partial_{[\mu}\xi_{\nu]}.
\ee
However, given that the metric is symmetric, this implies that the curvature scalar does not change
\be
R\rightarrow R,
\ee
{\em i.e.}~$R$ is invariant under projective transformations. Hence the Einstein--Hilbert action or any other action built from a function of $R$, such as the one used here, is projective invariant in metric-affine gravity.  However, the matter action is not generically projective invariant and this is the cause of the inconsistency of the field equations.

The conclusion that we have to draw is that when we want to consider a theory with a symmetric metric and an independent general connection, an action that depends only on the scalar curvature is not suitable. 
The ways to bypass this problem are then obvious: we have to drop one of the assumptions just listed. The first option is to abandon the requirement of a symmetric metric as in this case $R$, and consequently the gravitational action, would not be projectively invariant (see eq.~(\ref{projRicci})).  For the Einstein--Hilbert Lagrangian  this would lead to the well known Einstein--Straus theory \cite{schro}, and using an $f(R)$ Lagrangian would lead to a generalization of the it. This theory, even though it leads to fully consistent field equations,  is characterized by the fact that, in vacuum, neither non-metricity nor torsion vanish~\cite{schro}.  In particular this implies that torsion  in the Einstein--Strauss theory is not introduced by matter fields but it is intrinsic to gravity. Although logically possible, such an option does not seem very well motivated from a physical point of view, as one would more naturally expect any ``twirling" of spacetime to be somehow induced by the interaction with matter. We shall therefore not pursue this route any further. Instead we will consider the alternative solutions to our problem.

The second path towards a consistent theory is to modify the action by adding some extra terms. These terms should be chosen in such a way as to break projective invariance. There were proposals in this direction in the past, based on the study of an action linear in $R$ (see \cite{hehl2} and references therein). As an example we can mention the proposal of \cite{pap}: adding to the Lagrangian the term $g^{\mu\nu}\partial_\mu \Gamma^\sigma_{\phantom{a}[\nu\sigma]}$. Such a choice leads to a fully consistent theory and is mathematically very interesting. However, we find it difficult to physically motivate the presence of such a term in the gravitational action. Much more physically justified instead are corrections of the type $R^{\mu\nu}R_{\mu\nu}$, $R^{\alpha\beta\mu\nu}R_{\alpha\beta\mu\nu}$ etc. In fact, as we have already mentioned, such terms might very naturally be present in the gravitational action if we consider it as an effective, low energy, classical action coming from a more fundamental theory \cite{quant1,quant2,quant3,quant4,noji2,vassi}. We shall not discuss in detail such modifications here, since this goes beyond the scope of this paper; however, we shall make some comments. It is easy to verify, working for example with the simplest term $R^{\mu\nu}R_{\mu\nu}$, that such modifications will in general lead to consistent field equations. However, they will have the same attribute as the Einstein--Straus theory, {\em i.e.}~in vacuum, torsion will not generically vanish. One might imagine that a certain combination of higher order curvature invariants would lead to a theory with vanishing torsion in vacuum. To find such a theory would certainly be very interesting but is beyond the scope of the present investigation~\footnote{ One could even imagine to propose the absence of torsion in vacuum as a possible criterion in order to select a suitable combination of high energy (strong gravity) corrections to our $f(R)$ action.}. In conclusion this route generically leads to theories where again the presence of torsion seems an unmotivated complication rather than a physical feature. 

With no prescription about how to form a more general gravitational action which can lead to a physically attractive theory, we are left with only one alternative: to find a way of deriving consistent field equations with the action at hand. To understand how this is possible, we should re-examine the meaning of projective invariance. This is very similar to gauge invariance in Electromagnetism (EM). It tells us that the corresponding field, in this case the connections $\Gamma^\lambda_{\phantom{a}\mu\nu}$, can be determined from the field equations up to a projective transformation (eq.~(\ref{proj})). Breaking this invariance can therefore come by fixing some degrees of freedom of the field, similarly to a gauge fixing. The number of degrees of freedom which we need to fix is obviously the number of the components of the four-vector used for the transformation, {\em i.e.}~simply four. In practice this means that we should start by assuming that the connection is not the most general which one can construct, but it satisfies some constraints. Instead of placing an unphysical constraint on the action of the matter fields, as dictated by eqs.~(\ref{field22}) and (\ref{contr}), we can actually make a statement about spacetime properties. This is equivalent to saying that the matter fields can have all of the possible degrees of freedom but that the spacetime has some rigidity and cannot respond to some of them. (We shall come back to this point again later on. Let us just say that this is, for example, what happens in General Relativity when one assumes that there is no torsion and no non-metricity.)

We now have to choose the degrees of freedom of the connections that we need to fix. Since there are four of these, our procedure will be equivalent to fixing a four-vector. We can again let the studies of the Einstein--Hilbert action \cite{hehl2} lead the way. The proposal of Hehl et al.~\cite{hehl2} was to fix part of the non-metricity, namely the Weyl vector $Q_\mu$ (eq.~(\ref{weyl})). The easiest way to do this is by adding to the action a term containing a Lagrange multiplier $A^\mu$, which has the form
\be
S_{LM}=\int d^4 x \sqrt{-g} A^\mu Q_{\mu}.
\ee
 This way, one does not need to redo the variation of the rest of the action, but instead, only to evaluate the variation of the extra term. Varying with respect to the metric, the connections and $A$ respectively, we get the new field equations
\bea
\label{fieldQ1}
& &  f'(R) R_{(\mu\nu)}-\frac{1}{2}f(R)g_{\mu\nu}=\kappa T_{\mu\nu}+\frac{\kappa}{4\sqrt{-g}}\partial_\sigma(\sqrt{-g}A^\sigma)g_{\mu\nu},\\
\label{fieldQ2}
& &\frac{1}{\sqrt{-g}}\bigg[-\nabla_\lambda\left(\sqrt{-g}f'(R)g^{\mu\nu}\right)+\nabla_\sigma\left(\sqrt{-g}f'(R)g^{\mu\sigma}\right){\delta^\nu}_\lambda\bigg]+\nn\\& &\qquad\qquad+2f'(R)\left(g^{\mu\nu}S^{\phantom{ab}\sigma}_{\lambda\sigma}-g^{\mu\rho}S^{\phantom{ab}\sigma}_{\rho\sigma}{\delta^\nu}_\lambda+g^{\mu\sigma}S^{\phantom{ab}\nu}_{\sigma\lambda}\right)=
\nn\\& &\qquad\qquad\qquad\qquad\qquad\qquad\qquad\qquad=
\kappa\left(\Delta_{\lambda}^{\phantom{a}\mu\nu}-\frac{1}{4}{\delta^\mu}_\lambda A^\nu\right),\\
\label{Q0}
& & Q_\mu=0.
\eea
Taking the trace of eq.~(\ref{fieldQ2}) gives
\be
\label{DA}
A^\nu=\Delta_{\mu}^{\phantom{a}\mu\nu},
\ee
which is the consistency criterion, {\em i.e.}~it gives the value which we should choose for $A^\nu$ so that the equations are consistent. This procedure obviously works for $f(R)$ being a linear function as shown in \cite{hehl2}. However, we will demonstrate here that it cannot be used in any other case. 

 Let us consider the simple case of a matter action which does not depend on the connection. A good example could be a scalar field, as we have mentioned already, but as we shall see later there are also more conventional matter fields that have the same attribute. In this case eqs.~(\ref{fieldQ2}) and (\ref{DA}) give
\bea
\label{fieldQ22}
\frac{1}{\sqrt{-g}}\bigg[&-&\nabla_\lambda\left(\sqrt{-g}f'(R)g^{\mu\nu}\right)+\nabla_\sigma\left(\sqrt{-g}f'(R)g^{\mu\sigma}\right){\delta^\nu}_\lambda\bigg]+{}\nn\\ & &+2f'(R)\left(g^{\mu\nu}S^{\phantom{ab}\sigma}_{\lambda\sigma}-g^{\mu\rho}S^{\phantom{ab}\sigma}_{\rho\sigma}{\delta^\nu}_\lambda+g^{\mu\sigma}S^{\phantom{ab}\nu}_{\sigma\lambda}\right)=0.
\eea
One can consider separatelly the symmetric and antisymmetric parts of these equation with respect to the indices $\mu$ and $\nu$.
\bea
\label{fieldQ22sym}
& &\frac{1}{\sqrt{-g}}\bigg[-\nabla_\lambda\left(\sqrt{-g}f'(R)g^{\mu\nu}\right)+\nabla_\sigma\left(\sqrt{-g}f'(R)g^{\sigma(\mu}\right){\delta^{\nu)}}_\lambda\bigg]+{}\nn\\ 
& &\qquad\qquad +2f'(R)\left(g^{\mu\nu}S^{\phantom{ab}\sigma}_{\lambda\sigma}-g^{\rho(\mu}{\delta^{\nu)}}_\lambda S^{\phantom{ab}\sigma}_{\rho\sigma}+g^{\sigma(\mu}{S_{\sigma\lambda}}^{\nu)}\right)=0,\\
\label{fieldQ22ant}
& &  \frac{1}{\sqrt{-g}}\nabla_\sigma\left(\sqrt{-g}f'(R)g^{\sigma[\mu}\right){\delta^{\nu]}}_\lambda
+{}\nn\\ & &
\qquad\qquad+2f'(R)\left(-g^{\rho[\mu}{\delta^{\nu]}}_\lambda S^{\phantom{ab}\sigma}_{\rho\sigma}+g^{\sigma[\mu}{S_{\sigma\lambda}}^{\nu]} \right)=0.
\eea
A contraction between $\lambda$ and $\nu$ will straightforwardly lead to the equations
\bea
3\nabla_\sigma\left(\sqrt{-g}f'(R)g^{\sigma\mu}\right)&=& \, 4\sqrt{-g} f'(R)g^{\mu\rho}S^{\phantom{ab}\sigma}_{\rho\sigma},\\
\nabla_\sigma\left(\sqrt{-g}f'(R)g^{\sigma\mu}\right)&=& \, 0.
\eea
Combining these equations we can use them to write eqs.~(\ref{fieldQ22sym}) and (\ref{fieldQ22ant}) as
\bea
\label{fieldQ23sym}
&&-\frac{1}{\sqrt{-g}}\nabla_\lambda\left(\sqrt{-g}f'(R)g^{\mu\nu}\right)+2f'(R)g^{\sigma(\mu}{S_{\sigma\lambda}}^{\nu)}=0,\\
\label{fieldQ23ant}
&&g^{\sigma[\mu}{S_{\sigma\lambda}}^{\nu]}=0.
\eea
From eq.~(\ref{fieldQ23ant}), we get (if we use the metric to lower all of the indices)
\be
\label{not}
S_{\mu\lambda\nu} = S_{\nu\lambda\mu}.
\ee
This indicates that the Cartan torsion tensor must be symmetric with respect to the first and third index. However, by definition, it is also antisymmetric in the first two indices. Any third rank tensor with a symmetric and an antisymmetric pair of indices vanishes\footnote{Take the tensor $M_{\mu\nu\lambda}$ which is symmetric in its first and third index ($M_{\mu\nu\lambda}=M_{\lambda\nu\mu}$) and antisymmetric in the second and third index ($M_{\mu\nu\lambda}=-M_{\mu\lambda\nu}$). Exploiting these symmetries we can write
\bea
M_{\mu\nu\lambda}=M_{\lambda\nu\mu}=-M_{\lambda\mu\nu}=-M_{\nu\mu\lambda}=M_{\nu\lambda\mu}=M_{\mu\lambda\nu}=-M_{\mu\nu\lambda}\nonumber.
\eea
Therefore, $M_{\mu\nu\lambda}=0$.}. Thus
\be
S^{\phantom{ab}\nu}_{\sigma\lambda}=0,
\ee
torsion vanishes and we are left with the following equation
\be
\label{fieldQ24}
\nabla_\lambda\left(\sqrt{-g}f'(R)g^{\mu\nu}\right)=0.
\ee
This equation implies that one can define a metric $h_{\mu\nu}$ such that
\be
h_{\mu\nu}=f'(R)g_{\mu\nu},
\ee
which is covariantly conserved by the connections $\Gamma_{\phantom{a}\mu\nu}^\lambda$. This metric is, of course, symmetric since it is conformal to $g_{\mu\nu}$, and so the connections should by symmetric as well. Now notice the following: $h_{\mu\nu}$ has zero non-metricity by definition, leading to
\be
\nabla_\lambda h_{\mu\nu}=0.
\ee
A contraction with the metric will give
\be
4\frac{1}{f'(R)}\partial_\lambda f'(R)+g^{\mu\nu}f'(R)\nabla_\lambda g_{\mu\nu}=0
\ee
Now remember that eq.~(\ref{Q0}) forces the vanishing of the Weyl vector and that $Q_{\lambda}=g^{\mu\nu}\nabla_\lambda g_{\mu\nu}$. Therefore the above equation implies that
\be
\frac{1}{f'(R)}\partial_\lambda f'(R)=0,
\ee
{\em i.e.}~that $f'(R)$ is just a constant. If $f(R)$ is taken to be linear in $R$, everything is consistent, but this is not the case if one considers a more general $f(R)$ action. The reason for this is simply that part of the non-metricity in our theory is due to the form of the action. Therefore, constraining the non-metricity in any way turns out to be a constraint on the form of the Lagrangian itself which indicates that if we want to consider an action more general than the Einstein--Hilbert one, we should definitely avoid placing such kinds of constraint.

 The above exercise not only shows the lack of generality of the procedure adopted in \cite{hehl2} but also makes it clear that the four degrees of freedom which we have to fix are related to torsion. This implies that the torsionless version of the theory should be fully consistent without fixing any degrees of freedom. Let us now verify that. We can go back to the variation of the action in eq.~(\ref{varg3}) and force the connection to be symmetric. This gives
\bea
\label{varg3sym}
\delta S_G&=&\frac{1}{2\kappa}\int d^4 x\Bigg[\sqrt{-g}\left(f'(R) R_{(\mu\nu)}-\frac{1}{2}f(R)g_{\mu\nu}\right)\delta g^{\mu\nu}+{}\\& &+\left[-\nabla_\lambda\left(\sqrt{-g}f'(R)g^{\mu\nu}\right)+\nabla_\sigma\left(\sqrt{-g}f'(R)g^{\sigma(\mu}\right){\delta^{\nu)}}_\lambda\right]\delta\Gamma^\lambda_{\phantom{a}\mu\nu}\Bigg],\nn
\eea
and so the corresponding field equations are
\be
\label{field1sym}
f'(R) R_{(\mu\nu)}-\frac{1}{2}f(R)g_{\mu\nu}=\kappa T_{\mu\nu},\ee\be
\label{field2sym}
\frac{1}{\sqrt{-g}}\bigg[-\nabla_\lambda\left(\sqrt{-g}f'(R)g^{\mu\nu}\right)+\nabla_\sigma\left(\sqrt{-g}f'(R)g^{\sigma(\mu}\right){\delta^{\nu)}}_\lambda\bigg]=\kappa\Delta_{\lambda}^{\phantom{a}(\mu\nu)}.
\ee
where $\Delta_{\lambda}^{\phantom{a}\mu\nu}$ is also symmetrized due to the symmetry of the connection. One can verify easily that these equations are fully consistent. They are the field equations of $f(R)$ metric-affine gravity without torsion.

Turning back to our problem, we need to fix four degrees of freedom of the torsion tensor to make the torsionful version of the theory physically meaningful. A prescription has been given in \cite{sand} for a linear action and we shall see that it will work for our more general Lagrangian too. This prescription is to set the vector $S_\mu=S_{\sigma\mu}^{\phantom{ab}\sigma}$ equal to zero. Note that this does not mean that $\Gamma^{\phantom{ab}\sigma}_{\mu\sigma}$ should vanish but merely that $\Gamma^{\phantom{ab}\sigma}_{\mu\sigma}=\Gamma^{\phantom{ab}\sigma}_{\sigma\mu}$. We shall again use for this purpose a Lagrange multiplier $B^\mu$. The additional term in the action will be
\be
\label{lm2}
S_{LM}=\int d^4 x \sqrt{-g} B^\mu S_{\mu}.
\ee
It should be clear that the addition of this term does not imply that we are changing the action, since it is simply a mathematical trick to avoid doing the variation of the initial action under the assumption that $S_{\mu}=0$. The new field equations which we get from the variation with respect to the metric, the connections and $B^\mu$ are respectively
\bea
\label{field1t1}
& &f'(R) R_{(\mu\nu)}-\frac{1}{2}f(R)g_{\mu\nu}=\kappa T_{\mu\nu},\\
\label{field2t1}
& &\frac{1}{\sqrt{-g}}\bigg[-\nabla_\lambda\left(\sqrt{-g}f'(R)g^{\mu\nu}\right)+\nabla_\sigma\left(\sqrt{-g}f'(R)g^{\mu\sigma}\right){\delta^\nu}_\lambda\bigg]+{}\nn\\ & &\qquad\qquad+2f'(R)\left(g^{\mu\nu}S^{\phantom{ab}\sigma}_{\lambda\sigma}-g^{\mu\rho}S^{\phantom{ab}\sigma}_{\rho\sigma}{\delta^\nu}_\lambda+g^{\mu\sigma}S^{\phantom{ab}\nu}_{\sigma\lambda}\right)=\nn\\& &\qquad\qquad\qquad\qquad\qquad\qquad\qquad\qquad=\kappa(\Delta_{\lambda}^{\phantom{a}\mu\nu}-B^{[\mu}{\delta^{\nu]}}_{\lambda}),\\
& & S_{\mu\sigma}^{\phantom{ab}\sigma}=0.
\eea
Using the third equation we can simplify the second one to become
\bea
\frac{1}{\sqrt{-g}}\bigg[&-&\nabla_\lambda\left(\sqrt{-g}f'(R)g^{\mu\nu}\right)+\nabla_\sigma\left(\sqrt{-g}f'(R)g^{\mu\sigma}\right){\delta^\nu}_\lambda\bigg]+{}\nn\\ & &+2f'(R)g^{\mu\sigma}S^{\phantom{ab}\nu}_{\sigma\lambda}=\kappa(\Delta_{\lambda}^{\phantom{a}\mu\nu}-B^{[\nu}{\delta^{\mu]}}_{\lambda}).
\eea
Taking the trace over $\mu$ and $\lambda$ gives
\be
B^\mu=\frac{2}{3}\Delta_{\sigma}^{\phantom{a}\sigma\mu}.
\ee
Therefore the final form of the field equations is
\bea
\label{field1t}
& &f'(R) R_{(\mu\nu)}-\frac{1}{2}f(R)g_{\mu\nu}=\kappa T_{\mu\nu},\\
\label{field2t}
& &\frac{1}{\sqrt{-g}}\bigg[-\nabla_\lambda\left(\sqrt{-g}f'(R)g^{\mu\nu}\right)+\nabla_\sigma\left(\sqrt{-g}f'(R)g^{\mu\sigma}\right){\delta^\nu}_\lambda\bigg]+{}\nn\\ & &\qquad\qquad+2f'(R)g^{\mu\sigma}S^{\phantom{ab}\nu}_{\sigma\lambda}=\kappa(\Delta_{\lambda}^{\phantom{a}\mu\nu}-\frac{2}{3}\Delta_{\sigma}^{\phantom{a}\sigma[\nu}{\delta^{\mu]}}_{\lambda}),\\
\label{field3t}
& & S_{\mu\sigma}^{\phantom{ab}\sigma}=0.
\eea
These equations have no consistency problems and are the ones which we will be using for the rest of this paper.

So, in the end, we see that we can solve the inconsistency of the unconstrained field equations by imposing a certain rigidness on spacetime, in the sense that spacetime is allowed to twirl due to its interaction with the matter fields but only in a way that keeps $S_\mu=0$. This is not, of course, the most general case that one can think of, but as we demonstrated here, it is indeed the most general within the framework of $f(R)$ gravity.

We are now ready to further investigate the role of matter in determining the properties of spacetime. In particular we shall investigate the physical meaning of the hypermomentum $\Delta_{\lambda}^{\phantom{a}\mu\nu}$ and discuss specific examples of matter actions so as to gain a better understanding of the gravity-matter relation in the theories under scrutiny here.

\section{Matter actions}
\label{sec:MaAct}

In the previous section we derived the field equations for the gravitational field in the presence of matter. We considered both the case where torsion was allowed (eqs.~(\ref{field1t}), (\ref{field2t}) and (\ref{field3t})) and the torsionless version of the same theory (eqs.~(\ref{field1sym}) and (\ref{field2sym})). Observe that the first equation in both sets is the same, namely eqs.~(\ref{field1sym}) and (\ref{field1t}). The second one in each set is the one that has an explicit dependence on $\Delta_{\lambda}^{\phantom{a}\mu\nu}$, the quantity that it is derived when varying the matter action with respect to the connection, which has no analogue in General Relativity. 
We shall now consider separately more specific forms of the matter actions.

\subsection{Matter action independent of the connection}
\label{sec:noD}

Let us start by examining the simple case where the quantity $\Delta_{\lambda}^{\phantom{a}\mu\nu}$ is zero, {\em i.e.}~$S_M$ is independent of the connection.
In this case eq.~(\ref{field2t}) takes the form
\bea
\label{0}
\frac{1}{\sqrt{-g}}\bigg[&-&\nabla_\lambda\left(\sqrt{-g}f'(R)g^{\mu\nu}\right)+\nabla_\sigma\left(\sqrt{-g}f'(R)g^{\mu\sigma}\right){\delta^\nu}_\lambda\bigg]+{}\nn\\ & &+2f'(R)g^{\mu\sigma}S^{\phantom{ab}\nu}_{\sigma\lambda}=0.
\eea
Contracting the indices $\nu$ and $\lambda$ and using eq.~(\ref{field3t}) this gives
\be
\label{new1}
\nabla_\sigma\left(\sqrt{-g}f'(R)g^{\mu\sigma}\right)=0.
\ee
Using this result, eq.~(\ref{0}) takes the form
\bea
\label{anti0}
& &-\frac{1}{\sqrt{-g}}\nabla_\lambda\left(\sqrt{-g}f'(R)g^{\mu\nu}\right)+2f'(R)g^{\mu\sigma}S^{\phantom{ab}\nu}_{\sigma\lambda}=0.
\eea
Taking the antisymmetric part of this equation with respect to the indices $\mu$ and $\nu$  leads to 
\be
\label{notorsion}
g^{\sigma[\mu}{S_{\sigma\lambda}}^{\nu]}=0,
\ee
which can be written as
\be
\label{not2}
S_{\mu\lambda\nu} = S_{\nu\lambda\mu}.
\ee
As we explained previously, such a symmetry property leads to the vanishing of the torsion tensor when combined with its intrinsic antisymmetry with respect to the first two indices (see the footnote in the previous section). Thus
\be
S^{\phantom{ab}\nu}_{\sigma\lambda}=0.
\ee
The connection is now fully symmetric and the field equations are
\bea
\label{d01}
& &f'(R) R_{(\mu\nu)}-\frac{1}{2}f(R)g_{\mu\nu}=\kappa T_{\mu\nu},\\
\label{d02}
& &\nabla_\lambda\left(\sqrt{-g}f'(R)g^{\mu\nu}\right)=0.
\eea
Notice that these are the same equations that one derives for an {\it a priori} symmetric connection (see eqs.~(\ref{field1sym}) and (\ref{field2sym})); {\em i.e.}~for $\Delta_{\lambda}^{\phantom{a}\mu\nu}=0$, the torsionless and the torsionful versions of the theory coincide. Eq.~(\ref{d02}) implies that one can define a metric $h_{\mu\nu}$ such that
\be
h_{\mu\nu}=f'(R)g_{\mu\nu},
\ee
which is covariantly conserved by the connections $\Gamma_{\phantom{a}\mu\nu}^\lambda$. This metric is, of course, symmetric since it is conformal to $g_{\mu\nu}$, and so the connections should be symmetric as well.
In other words, it has been shown that $\Delta_{\lambda}^{\phantom{a}\mu\nu}=0$ leads to a symmetric connection, which means that there is no torsion when the matter action does not depend on the connection. This is an important aspect of this class of metric-affine theories of gravity. It shows that {\it metric-affine  $f(R)$ gravity allows the presence of torsion but does not force it}. {\it Torsion is merely introduced by specific forms of matter}; those for which the matter action has a dependence on the connections. Therefore, as ``matter tells spacetime how to curve'', matter will also tell spacetime how to twirl. Notice also that the non-metricity does not vanish. This is because, as we also saw previously, part of the non-metricity is introduced by the form of the Lagrangian, {\em i.e.}~$f(R)$ actions lead generically to theories with intrinsic non-metricity.

It is interesting to note again the special nature of the particular case in which the $f(R)$ Lagrangian is actually linear in $R$, {\em i.e.}
\be
f(R)=R-2\Lambda.
\ee
Then eq.~(\ref{d01}) gives
\be
R_{(\mu\nu)}-\frac{1}{2}R g_{\mu\nu}+\Lambda g_{\mu\nu}=\kappa T_{\mu\nu},
\ee
and eq.~(\ref{d02}) gives
\be
\Gamma_{\phantom{a}\mu\nu}^\lambda=\{_{\phantom{a}\mu\nu}^\lambda\},
\ee
{\em i.e.}~the $\Gamma$'s turn out to be the Levi--Civita connections of the metric and so the theory actually reduces to standard GR which, from this point of view, can now be considered as a sub-case of a metric-affine theory. Finally, let us also mention that in vacuum both $\Delta_{\lambda}^{\phantom{a}\mu\nu}$ and $T_{\mu\nu}$ are equal to zero, so that the field equations reduce straightforwardly, as expected, to eqs.~(\ref{e4}) and (\ref{e5}) of section \ref{nonlinear}.

\subsection{Matter action dependent on the connection}
\label{yesD}

Having studied the field equation for matter fields whose Lagrangian does not depend on the connections ($\Delta_{\lambda}^{\phantom{a}\mu\nu}=0$), let us now proceed to the more general case where there is such a dependence ($\Delta_{\lambda}^{\phantom{a}\mu\nu}\neq 0$). We can find two interesting sub-cases here. These are when $\Delta_{\lambda}^{\phantom{a}\mu\nu}$ is either fully symmetric or fully antisymmetric in the indices $\mu$ and $\nu$. As before, the equation under investigation will be eq.~(\ref{field2t}). We shall split it here into its symmetric and antisymmetric parts in the indices $\mu$ and $\nu$:
\bea
\label{2sym}
& &\frac{1}{\sqrt{-g}}\bigg[-\nabla_\lambda\left(\sqrt{-g}f'(R)g^{\mu\nu}\right)+\nabla_\sigma\left(\sqrt{-g}f'(R)g^{\sigma(\mu}\right){\delta^{\nu)}}_\lambda\bigg]+{}\nn\\ & &\qquad\qquad+2f'(R)g^{\sigma(\mu}{S_{\sigma\lambda}}^{\nu)}=\kappa\Delta_{\lambda}^{\phantom{a}(\mu\nu)},\\
\label{2anti}
& &\frac{1}{\sqrt{-g}}\nabla_\sigma\left(\sqrt{-g}f'(R)g^{\sigma[\mu}\right){\delta^{\nu]}}_\lambda+2f'(R)g^{\sigma[\mu}{S_{\sigma\lambda}}^{\nu]}=\nn\\
& &\qquad\qquad=\kappa(\Delta_{\lambda}^{\phantom{a}[\mu\nu]}-\frac{2}{3}\Delta_{\sigma}^{\phantom{a}\sigma[\nu}{\delta^{\mu]}}_{\lambda}).
\eea
Let us assume now that
\be
\label{dsym}
\Delta_{\lambda}^{\phantom{a}[\mu\nu]}=0,
\ee
and take the trace of any of the above equations. This leads to
\be
3\nabla_\sigma\left(\sqrt{-g}f'(R)g^{\sigma\mu}\right)=2\sqrt{-g}\kappa\Delta_{\sigma}^{\phantom{a}\sigma\mu}.
\ee
Using this and eq.~(\ref{dsym}), eq.~(\ref{2anti}) takes the form
\be
g^{\sigma[\mu}S^{\phantom{ab}\nu]}_{\sigma\lambda}=0,
\ee
which is the same as eq.~(\ref{notorsion}) which we have shown leads to
\be
S^{\phantom{ab}\nu}_{\sigma\lambda}=0.
\ee
Then, once again, the torsion tensor vanishes and we drop to the system of equation
\be
f'(R) R_{(\mu\nu)}-\frac{1}{2}f(R)g_{\mu\nu}=\kappa T_{\mu\nu},\ee\be
\frac{1}{\sqrt{-g}}\bigg[-\nabla_\lambda\left(\sqrt{-g}f'(R)g^{\mu\nu}\right)+\nabla_\sigma\left(\sqrt{-g}f'(R)g^{\sigma(\mu}\right){\delta^{\nu)}}_\lambda\bigg]=\kappa\Delta_{\lambda}^{\phantom{a}(\mu\nu)}.
\ee
which are the same as eqs.~(\ref{field1sym}) and (\ref{field2sym}) {\em i.e.}~the equations for the torsionless version of the theory. This indicates that any torsion is actually introduced by the antisymmetric part of $\Delta_{\lambda}^{\phantom{a}\mu\nu}$.

We can now examine the opposite case where it is the symmetric part of $\Delta_{\lambda}^{\phantom{a}\mu\nu}$ that vanishes. Then
\be
\label{danti}
\Delta_{\lambda}^{\phantom{a}(\mu\nu)}=0,
\ee
and taking the trace of either eq.~(\ref{2sym}) or eq.~(\ref{2anti}) straightforwardly gives
\be
\nabla_\sigma\left(\sqrt{-g}f'(R)g^{\sigma\mu}\right)=0.
\ee
Therefore, eqs.~(\ref{2sym}) and (\ref{2anti}) take the form
\bea
\label{2symanti}
& &-\frac{1}{\sqrt{-g}}\nabla_\lambda\left(\sqrt{-g}f'(R)g^{\mu\nu}\right)+2f'(R)g^{\sigma(\mu}{S_{\sigma\lambda}}^{\nu)}=0,\\
\label{2antianti}
& &2f'(R)g^{\sigma[\mu}{S_{\sigma\lambda}}^{\nu]}=\kappa(\Delta_{\lambda}^{\phantom{a}[\mu\nu]}-\frac{2}{3}\Delta_{\sigma}^{\phantom{a}\sigma[\nu}{\delta^{\mu]}}_{\lambda}).
\eea
 Taking into account the general expression for the covariant derivative of a tensor density
\be
\nabla_\lambda(\sqrt{-g} J^{\alpha\dots}_{\:\:\beta\dots})=\sqrt{-g}\nabla_\lambda( J^{\alpha\dots}_{\:\:\beta\dots})-\sqrt{-g}\Gamma^{\sigma}_{\phantom{a}\sigma\lambda}J^{\alpha\dots}_{\:\:\beta\dots},
\ee
and the fact that $\Gamma^{\sigma}_{\phantom{a}\sigma\lambda}=\Gamma^{\sigma}_{\phantom{a}\lambda\sigma}$ by eq.~(\ref{field3t}), one can easily show that the eq.~(\ref{2symanti}) can be written as
\be
\bar{\nabla}_\lambda\left(\sqrt{-g}f'(R)g^{\mu\nu}\right)=0,
\ee
where $\bar{\nabla}_\lambda$ denotes the covariant derivative defined with the symmetric part of the connection. This equation tells us that, as before, we can define a symmetric metric
\be
h_{\mu\nu}=f'(R)g_{\mu\nu},
\ee
which is now covariantly conserved by the symmetric part of  connections, $\Gamma_{\phantom{a}(\mu\nu)}^\lambda$. If $f(R)$ is linear in $R$, $h_{\mu\nu}$ and $g_{\mu\nu}$ coincide, of course. Additionally, eq.~(\ref{2antianti}) shows that the torsion is fully introduced by the matter fields. Therefore we can conclude that when $\Delta_{\lambda}^{\phantom{a}\mu\nu}$ is fully antisymmetric, there is torsion, but the only non-metricity present is that introduced by the form of the  gravitational Lagrangian, {\em i.e.}~matter introduces no extra non-metricity.

 We can now safely conclude that, in the metric-affine framework discussed here, matter can induce both non-metricity and torsion: the symmetric part of $\Delta_{\lambda}^{\phantom{a}\mu\nu}$ introduces non-metricity, the antisymmetric part is instead responsible for introducing torsion. While some non-metricity is generically induced also by the $f(R)$ Lagrangian (with the relevant exception of the linear case), torsion is only a  product of the presence of matter.

\subsection{Specific matter fields}
\label{spmfields}

 Having studied the implication of a vanishing or non vanishing $\Delta_{\lambda}^{\phantom{a}\mu\nu}$, we now want to discuss these properties in terms of specific fields.
Since $\Delta_{\lambda}^{\phantom{a}\mu\nu}$ is the result of the variation of the matter action with respect to the connection, we will need the matter actions of the fields in curved spacetime for this purpose. In purely metric theories one knows that any covariant equation, and hence also the action, can be written in a local inertial frame by assuming that the metric is flat and the connections vanish, turning the covariant derivatives into partial ones. Therefore, one can expect that the inverse procedure, which is called the minimal coupling principle, should hold as well and can be used to provide us with the matter action in curved spacetime starting from its expression in a local inertial frame. This expectation is based on the following conjecture: {\it The components of the gravitational field should be used in the matter action on a necessity basis}. The root of this conjecture can be traced to requiring minimal coupling between the gravitational field and the matter fields (hence the name ``minimal coupling principle''). In General Relativity this conjecture 
can be stated for practical purposes in the following form: {\it the metric should be used in the matter action only for contracting indices  and constructing the terms that need to be added in order to write a viable covariant matter action}. This implies that the connections should appear in this action only inside covariant derivatives and never alone which is, of course, perfectly reasonable since, first of all, they are not independent fields and, secondly, they are not tensors themselves and so they have no place in a covariant expression. At the same time, other terms that would vanish in flat spacetime, like, for example, contractions of the curvature tensor with the fields or their derivatives, should be avoided. 

The previous statements are not applicable in metric affine gravity for several reasons: the connections now are independent fields and, what is more, if they are not symmetric, there is a tensor that one can construct via their linear combination: the Cartan torsion tensor.  Additionally, going to some local inertial frame in metric-affine gravity is a two-step procedure in which one has to separately impose that the metric is flat and that the connections vanish. However, the critical point is that when inverting this procedure one should keep in mind that there might be dependences from the connections in the equations other than those in the covariant derivatives. {\it The standard  minimal coupling principle will therefore not, in general, give the correct answer in metric-affine gravity theories.}

The above discussion can be well understood through a simple example, using the electromagnetic field. 
 In order to compute the hypermomentum  $\Delta_{\lambda}^{\phantom{a}\mu\nu}$ of the electromagnetic field, we need to start from the action
\be
\label{elaction}
S_{EM}=-\frac{1}{4}\int d^4 x \sqrt{-g} F^{\mu\nu}F_{\mu\nu},
\ee
where $F^{\mu\nu}$ is the electromagnetic field tensor. As we know, in the absence of gravity this tensor is defined as
\be
\label{elf}
F_{\mu\nu}\equiv \partial_\mu A_{\nu}-\partial_\nu A_{\mu},
\ee
where $A_{\mu}$ is the electromagnetic four-potential. If we naively followed the minimal coupling principle and simply replaced the partial derivatives with covariant ones,
 the definition of the the electromagnetic field tensor would take the form:
\be
\label{wrong}
F_{\mu\nu}\equiv \nabla_\mu A_{\nu}-\nabla_\nu A_{\mu}=\partial_\mu A_{\nu}-\partial_\nu A_{\mu}-2\,\Gamma^\sigma_{\phantom{a}[\mu\nu]}A_\sigma,
\ee
and one can easily verify that the electromagnetic field tensor would then no longer be gauge invariant, {\em i.e.}~invariant under redefinition of the four potential of the form $A_{\mu}\rightarrow A_{\mu}+\partial_\mu \phi$, where $\phi$ is a scalar quantity. Gauge invariance, however, is a critical aspect of the electromagnetic field since it is related to the conservation of charge and the fact that the electric and magnetic fields are actually measurable quantities. Therefore breaking gauge invariance cannot lead to a viable theory.  One could assume that the problem lies in the fact that the connection is not symmetric, {\em i.e.}~torsion is allowed, since it is the antisymmetric part of the connection that prevents gauge invariance of eq.~(\ref{wrong}), and hence it might seem that standard elctromagnetism is incompatible with torsion. This explanation was given for example in~\cite{maj} (see also references therein for other discussions following the same line). We do not agree with such an approach: As we said, the problem is actually much simpler but also more fundamental and lies in the assumption that the minimal coupling principle still holds in metric-affine gravity.

In order to demonstrate this point, let us turn our attention to the definition of the electromagnetic field tensor in the language of differential forms. This is
\be
\label{defdf}
{\bf F}\equiv{\bf d}{\bf A},
\ee
where ${\bf d}$ is the standard exterior derivative \cite{grav}. Remember that the exterior derivative is related to the Gauss theorem which allows us to go from an integral over the volume to an integral over the boundary surface of this volume. Now notice that the volume element has no dependence on the connection and is the same as that of general relativity, $\sqrt{-g}\, d^4 x$. This implies that the definition of the exterior derivative should remain unchanged when expressed in terms of partial derivatives. Partial derivatives on the other hand are defined in the same way in this theory as in general relativity. Therefore, from the definition (\ref{defdf}) we understand that $F_{\mu\nu}$ should be given in terms of the partial derivatives by the following equation
\be
\label{elf2}
F_{\mu\nu}\equiv {\rm d}A=\partial_\mu A_{\nu}-\partial_\nu A_{\mu},
\ee
which is the same as eq.~(\ref{elf}) and respects gauge invariance. The expression in terms of partial derivatives is  not covariant but can easily be written in a covariant form:
\bea
\label{right}
F_{\mu\nu}\equiv \partial_\mu A_{\nu}-\partial_\nu A_{\mu}&=&\nabla_\mu A_{\nu}-\nabla_\nu A_{\mu}+2\,\Gamma^\sigma_{\phantom{a}[\mu\nu]}A_\sigma=\nn\\&=&\nabla_\mu A_{\nu}-\nabla_\nu A_{\mu}+2\,S^{\phantom{ab}\sigma}_{\mu\nu}A_\sigma.
\eea

It is obvious now that the minimal coupling principle was leading us to the wrong expression, causing a series of misconceptions. However, we are still in need of a prescription that will allow us to derive the matter actions in curved spacetime. Notice that if we require gravity and matter to be minimally coupled, then the physical basis of the conjecture  that {\it the components of the gravitational field should be used in the matter action on a necessity basis}  still holds, since its validity is not related to any of the assumptions of General Relativity. Thus, we can use it to express a metric-affine minimal coupling principle: {\it The metric should be used in the matter action only for contracting indices and the connection should be used in order to construct the extra terms that we must to add in order to write a viable covariant matter action}. The analogy with the statement used in General Relativity is obvious, and differences lie in the different character of the connections in the two theories. One can easily verify that the matter action of the electromagnetic field which we derived earlier can be straightforwardly constructed using this metric--affine minimal coupling principle. 

We would like to stress once more that both the metric--affine minimal coupling principle presented above as well as the standard one, are based on the requirement that the gravitational field should be minimally coupled to the matter.
One could of course choose to construct a theory without such a requirement, and allow non-minimal coupling~\footnote{Notice that if one considers the possible actions for classical gravity as effective ones --- obtained as the low energy limit of some more fundamental high energy theory --- then it is natural to imagine that the form of the coupling (minimal or some specific type of non-minimal) might cease to be a free choice (see {\em e.g.}~chapter 7 of \cite{faraoni} for an enlightening discussion). However, one could still expect that non-minimal coupling terms will be suppressed at low energies by appropriate powers of the scale associated with the fundamental theory (Planck scale, string scale, etc.) and in this sense the use of a minimal coupling principle at low energies could be justified.}.
This can be done both in metric-affine gravity and in General Relativity. Clearly, in metric-affine gravity one has more options when it comes to non-minimal coupling, since besides curvature terms, also terms containing the Cartan torsion tensor can be used. However, it is easy to see that the number of viable coupling terms is strongly reduced by the symmetry of the metric (which also implies symmetry of the stress-energy tensor) and by the constraints of the theory, {\em e.g.}~the vanishing of the trace of  $S^{\phantom{ab}\sigma}_{\mu\nu}$ when considering $f(R)$ actions.

Allowing non-minimal coupling between gravity and matter in a gravitational theory drastically changes the corresponding phenomenology and there might be interesting prospects for such attempts in metric-affine gravity. For the rest of this paper, however, we will continue to assume minimal coupling between gravity and matter, since this is the most conventional option.

Let us now return to the electromagnetic field. Now that we have  a suitable expression for the electromagnetic field tensor we can proceed to derive the field equations for electrovacuum. Notice that $F_{\mu\nu}$ has no real dependence on the connections and so we can straightforwardly write
\be
\label{Del}
\Delta_{\lambda}^{\phantom{a}\mu\nu}=0.
\ee
The stress-energy tensor $T_{\mu\nu}$ can be evaluated using eq.~(\ref{set}) and has the standard form
\be
T_{\mu\nu}=F_\mu^{\phantom{a}\sigma}F_{\sigma\nu}-\frac{1}{4}g_{\mu\nu}F^{\alpha\beta}F_{\alpha\beta}.
\ee
With the use of eqs.~(\ref{d01}) and(\ref{d02}) we can write the field equations:
\bea
\label{field1el}
& &f'(R) R_{(\mu\nu)}-\frac{1}{2}f(R)g_{\mu\nu}=\kappa F_\mu^{\phantom{a}\sigma}F_{\sigma\nu}-\frac{\kappa}{4}g_{\mu\nu}F^{\alpha\beta}F_{\alpha\beta},\\
\label{field2el}
& &\nabla_\lambda\left(\sqrt{-g}f'(R)g^{\mu\nu}\right)=0.
\eea
We can use, however, the fact that the stress-energy tensor of the electromagnetic field is traceless. If we take the trace of eq.~(\ref{field1el}) we get
\be
f'(R)R-2 f(R)=0,
\ee
which, as we discussed previously for the vacuum case, is an algebraic equation in $R$ once $f(R)$ has been specified. Solving it will give a number of roots (see also the discussion after eq.~(\ref{sol}))
\be
R=c_i,\quad i=1,2,\dots
\ee
and $f(c_i)$ and $f'(c_i)$ will be constants. Thus, eq.~(\ref{field2el}) implies that the metric is covariantly conserved by the covariant derivative defined using the connection and so
\be
\Gamma_{\phantom{a}\mu\nu}^\lambda=\{_{\phantom{a}\mu\nu}^\lambda\},
\ee
and we are left with the following field equation:
\be
\label{field1eleh}
R_{\mu\nu}-\frac{1}{4}c_i g_{\mu\nu}=\kappa' F_\mu^{\phantom{a}\sigma}F_{\sigma\nu}-\frac{\kappa'}{4}g_{\mu\nu}F^{\alpha\beta}F_{\alpha\beta},
\ee
which is the Einstein equation for electrovacuum with a cosmological constant and a modified ``coupling constant'' $\kappa'=\kappa/f'(c_i)$. The rescaling of $\kappa$ should not mislead us into thinking that either the gravitational constant, $G$, or the fine structure constant, $\alpha$, change in any way. It just affects the strength of the ``coupling'' between gravity and the electromagnetic field, {\em i.e.}~how much curvature is induced per unit energy of the electromagnetic field. The values of the cosmological constant and $\kappa'$ depend on the functional form of $f(R)$ and therefore they are fixed once one selects an action. For example, $f(R)=R$ or $f(R)=a R^2+R$, both lead to $c_i=0$ and $\kappa'=\kappa$, and the theory will be indistinguishable from General Relativity. For more general forms of $f(R)$, the theory is still formally equivalent to General Relativity but notice that the modification of $\kappa$ should, at least theoretically, be subject to experiment. If such an experiment is technically possible it might help us place bounds on the form of the action.

As already mentioned, a vanishing $\Delta_{\lambda}^{\phantom{a}\mu\nu}$ implies that there is no dependence of the matter action on the connections, or equivalently on the covariant derivative. As we just saw, the elactromagnetic field, and consequently any other gauge field, has this attribute. The same is true for a scalar field, as the covariant derivatives of a scalar are reduced to partial derivatives. Therefore, neither of these fields will introduce torsion or extra non-metricity. For the electromagnetic field specifically, the fact that the trace of its stress energy tensor is zero leads to the Einstein field equations, since the non-metricity introduced by the form of the Lagrangian has to vanish as well. For the scalar field, whose stress energy tensor does not have a vanishing trace, this will not happen. The field equations can be derived straightforwardly by replacing the usual stress energy tensor of a scalar field in eqs.~(\ref{d01}) and (\ref{d02}).

 Let us now turn to matter fields for which $\Delta_{\lambda}^{\phantom{a}\mu\nu}$ does not vanish. In principle any massive vector field or tensor field should have an action with an explicit dependence on the connection leading to non vanishing $\Delta_{\lambda}^{\phantom{a}\mu\nu}$. A typical example would be the Dirac field. The Dirac Lagrangian has an explicit dependence on the covariant derivative, and therefore an explicit dependence on the connections. Additionally, there are no viability criteria, unlike in the case of the electromagnetic field, that will force us to include extra terms proportional to the Cartan torsion tensor which will cancel out the presence of the antisymmetric part of the connection. Therefore, the procedure for  deriving the matter action is straightforward (see \cite{hehlrev} for the full form of the action \footnote{Note that the result of \cite{hehlrev} is for a theory that has, by definition, vanishing non-metricity ($U_4$ theory). However the form of the  matter action is the same once the proper covariant derivative is used. For discussions about the matter actions in theories with torsion see also \cite{benn,carfie}. Note that, even though the results obtained here are in complete agreement with the ones presented in those works, in many cases the reasoning differs, since there is no attempt there to formulate a metric-affine minimal coupling principle. The standard principle is there used in cases where it provides the correct results while it is noted that it does not apply to specific cases, such as the electromagnetic field. For each of these cases individual arguments are used in order to derive the matter action in curved spacetime. The underlying physics in the two approaches is the same, but we believe that the idea of a metric-affine minimal coupling principle is an essential concept since, besides its elegance and analogy with the standard minimal coupling principle, it leaves no room for exceptions.}). We can infer from the above that a Dirac field will potentially introduce both torsion and non-metricity. Note that the fields which cannot introduce torsion will also not ``feel'' it, since they are not coupled to the Cartan tensor, and so, photons or scalar particles will not be affected by torsion, even if other matter fields produce it.

It is also interesting to study matter configurations in which matter is treated macroscopically, the most common being that of  a perfect fluid. Let us here consider separately the cases where torsion is allowed in the theory and where it is not included. In the latter case the consideration of the perfect fluid is identical to standard GR. Since the matter action can be described by two scalars, the energy density and the velocity potential (see for example \cite{scha,stone}), the action has no dependence on the covariant derivative and so  $\Delta_{\lambda}^{\phantom{a}\mu\nu}$ will vanish. When torsion is allowed, there two distinct cases depending on the microscopic properties of the fluid. If a perfect fluid is used to effectively describe particles whose corresponding field description does not introduce torsion then no difference from the previous case arises. If however the fluid is composed of this kind of particles, then their spin has to be taken into account (see \cite{hehlrev} and references therein). There will however be an averaging over volume of the quantities describing the matter, and if one assumes that the spin is randomly oriented and not polarized, then it should average to zero. This description can be applied in physical situations such as graviational collapse or cosmology. The fact that the expectation value of the spin will be zero will lead to a vanishing expectation value for the torsion tensor. However, fluctuations around the expectation value will affect the geometry, leading to corrections to the field equations, which will depend on the energy density of the specific species of particle. Since the 
torsion tensor is coupled to the hypermomentum thought the gravitational constant (eq.~(\ref{field2t})), the effect of these fluctuations will be suppressed by a Planck mass square. Therefore we can conclude that for cosmology, and especially for late times where the energy density is small, the standard perfect fluid description might serve as an adequate approximation. 

It is remarkable that the two matter descriptions most commonly used in cosmology, the perfect fluid and the scalar field, lead to a vanishing  $\Delta_{\lambda}^{\phantom{a}\mu\nu}$ for a symmetric connection. It is also noticeable that in our framework, even if torsion is allowed, the results remain unchaged apart from small corrections for the perfect fluid case, which should be negligible. It would be interesting to consider also the case of a an imperfect fluid ({\it i.e.}~to allow also viscosity, heat flow, etc.), which is certainly relevant for some observationally interesting systems in relativistic astrophysics. 
As in the case of a perfect fluid, if we consider particles with a  
spin and allow torsion, the standard imperfect fluid description will  
not be exact. Note however, that even in the simpler case of {\em a priori} symmetric connections, we do not expect the matter action to be independent of such connections (in contrast with the perfect fluid case). This could lead to a non-vanishing $\Delta_\lambda^{\phantom{a}\mu\nu}$ and consequently to some non-metricity, which might lead to interesting deviations from GR results.

\subsection{Discussion}
\label{disc}

As already mentioned, since torsion is absent in vacuum or in some specific matter configurations but present in all other cases, we can infer that it is actually introduced by matter. By considering for which kind of fields torsion vanishes and for which it does not, we can arrive at a very interesting conclusion. Torsion is zero in vacuum and in the presence of a scalar field or the electromagnetic field. It does not necessarily vanish, however, in the presence of a Dirac field or other vector and tensor fields. This shows a correspondence between torsion and the presence of fields that describe particles with spin. We are, therefore, led to the idea that particles with spin seem to be the sources of torsion. Of course a photon, the particle associated with the electromagnetic field, is a spin one particle. However, in Quantum Field Theory a photon is not really characterized by its spin but actually by its helicity. It is remarkable that this exceptional nature of the photon seems to be present also here, since the electromagnetic field is unable to introduce torsion.

We have also discussed the case where matter is treated macroscopically. As already mentioned, a perfect fluid cannot introduce any extra non-metricity for a symmetric connection. When torsion is allowed,  the concept of a perfect fluid has to be generalized if one wants to include particles with spin,  but also in this case only small contributions to torsion will be introduced, which will be negligible in most cases. On the other hand, for both symmetric and general connections, we expect deviations from GR when an imperfect fluid is considered. This is a very important aspect, since in many applications, like cosmology and astrophysics, matter is treated as a perfect fluid.  Moreover, many of the experimental tests passed by GR are related to either vacuum or to environments where matter can more or less be accurately described as a perfect fluid. This means that a metric-affine theory could be in total accordance with these tests when the Einstein--Hilbert action and possibly many of its extensions are used. However, the possible relevance of imperfect fluid matter in some yet to be accurately observed astrophysical systems (like accretion flows or compact objects \cite{john}) leaves open the possibility for future discrimination between the class of theories discussed here and standard GR.

 Let us conclude this section with a comment on the cosmological aspects of metric-affine $f(R)$ theories of gravity. As mentioned before there are numerous studies in this subject in $f(R)$ gravity in the Palatini formalism, where the matter action is assumed to be independent on the connections. Even though, as we have argued, such an {\em a priori} assumption is unphysical in the context of metric-affine gravity, the results of these papers (see {\em e.g.}~\cite{sot1,vollick,meng3,meng,sot2,mw,mw2}) are perfectly valid also in this context due to the fact that the connections are assumed to be symmetric and all of the calculations are performed using a perfect fluid description for the matter. However it should be by now clear that, in performing similar studies with metric-affine gravity, the independence of the matter action from the connections cannot be assumed {\em a priori} and even when verified, it can be at most due to the specific kind of matter considered, not to a general feature of the theory. 

This discussion is also relevant to a recent debate between Flanagan and Vollick \cite{flan,voll1,flan2,voll2}. Flanagan showed in \cite{flan} that the metric-affine version of $1/R$ gravity presented in \cite{vollick} is in conflict with the standard model of particle physics. The conclusions presented in~\cite{flan} were obtained by considering Dirac fermions as the matter field. However, it was soon observed, in \cite{voll2}, that the action for such a Dirac field was built in~\cite{flan} using the Levi--Civita connections of the metric and  not the connections used for constructing the gravitational action of the theory. Additionally, it was shown that once this assumption is dropped and one uses, in the matter action as well, the covariant derivative related to the connection used in the gravitational action, the results of \cite{flan} no longer hold.

 On the other hand, it has to be stressed that the model of \cite{vollick} did assume {\em a priori} a generic matter action which was independent of the true connections. In this sense Flanagan's approach in~\cite{flan} was in accordance with the philosophy of the work it intended to criticize. Using the results presented here, we can say the following: models with $1/R$ terms in the gravitational action seem to be in conflict with the standard model of particle physics as shown in \cite{flan} if no dependence of any matter action on the true connections is assumed. However, if one does not make this assumption {\em a priori}, but uses the full version of the theory described here, then this conflict is removed. Moreover, since in these cosmological models matter is always treated as a perfect fluid, their most interesting phenomenological implications will remain unaffected.

\section{Conclusions}
\label{conc}
We have studied here metric-affine $f(R)$ theories of gravity, {\em i.e.}~theories of gravity where the gravitational action can be a general function of the scalar curvature and the metric and the affine connections are considered as independent quantities. We have derived the gravitational field equations both in vacuum and in the presence of matter fields. In our effort to allow the connections to be in general non-symmetric to consider also torsion, we saw that serious difficulties occur, coming from the fact that for an arbitrary connection the gravitational Lagrangian of $f(R)$ theories turns out to be projective invariant, unlike the matter Lagrangian. In order to overcome this problem without resorting to another form of the gravitational action, one has to fix four degrees of freedom of the torsion, and this was the approach followed here. 

It was shown that when the variation of the matter action leads to a tensor symmetric in its last two indices, then torsion vanishes. When the same tensor is antisymmetric, matter introduces only torsion and not non-metricity. Matter fields whose matter action is independent of the connection, such as a scalar field, cannot introduce either torsion or non-metricity. Additionally, the whole theory reduces to General Relativity with a cosmological constant, either in vacuum or electrovacuum. The study of the electromagnetic field turned out very helpful, since it demonstrated that the minimal coupling principle  does not hold in metric-affine gravity. However, as we showed, one can express a metric-affine minimal coupling principle, based on the spirit of minimal coupling between gravity and matter. Finally, our study revealed a connection between the presence of torsion and the presence of fields describing particles with non-zero spin. In the absence of such particles torsion becomes zero leading to the idea that it is actually introduced by these specific forms of matter. We find this picture very appealing and physical since it seems to indicate that in metric-affine gravity {\it 
 as matter tells spacetime how to curve, matter will also tell spacetime how to twirl}.

We have also considered macroscopic descriptions of matter showing that a perfect fluid will not introduce non-metricity or torsion, since its matter action has no explicit dependence on the connections. 
We briefly discussed the implications of this in recent applications of metric affine gravity to cosmology. There are many physical systems, however, where matter cannot necessarily be described accurately enough by a perfect fluid. In these cases one would expect to see a deviation from the standard behaviour of General Relativity. Even starting with the standard Einstein--Hilbert action, torsion and non-metricity should affect the dynamics and make them deviate from the standard ones. This deviation could persist even in a nearly-Newtonian regime. It would though, disappear in vacuum or in an environment where matter can be described sufficiently well by a perfect fluid, the circumstances in which most of the weak field gravitational tests are held. However, it might be very interesting to study this in the context of galactic dynamics since in this case the effects may be important and even make some contribution in relation to the unexpected behavior of the rotational curves. Of course, until a thorough and quantitative study is performed, all of the above remain on the level of speculations, even though they seem qualitatively interesting. We hope to address this problem in future work. 

It is important to notice that our attempt to include torsion showed that this is not possible in the context of $f(R)$ gravity unless one fixes some degrees of freedom of the connection as mentioned earlier. The other possibility that was discussed here was to modify the action by adding some higher order curvature invariant. As we said, it is very difficult to find a prescription for an action of this form that will lead to a physically meaningful theory of gravitation with torsion since the simple case will have unwanted attributes. This is the reason why we did not pursue this here. Note however, that we already know that rotating test particles do not follow geodesics. Therefore, it would be reasonable to assume that, since macroscopic angular momentum interacts with the geometry, intrinsic angular momentum (spin) should interact as well. This property should become more important at small scales or high energies. Therefore, it seems remarkable that an attempt to include torsion and at the same time avoid placing {\it a priori} constraints on the connection, leads to the conclusion that the action should be supplemented with higher order curvature invariants, which is in total agreement with the predictions coming from quantum corrections, String theory and M-theory. 

To conclude, we would like to say that metric-affine theories of gravity seem to constitute important and interesting candidates for a modified theory of gravity. They seem to reduce to General Relativity, or a theory very close to that, in most of the cases relevant to known experimental tests and at the same time are phenomenologically much richer. This may help to address some of the puzzles of physics related to gravity.

\section*{Acknowledgments}

The authors would like to thank John Miller and Sebastiano Sonego for illuminating discussions. T.~P.~S.~would also like to thank Salvatore Capozziello and Mauro Francaviglia for interesting discussion about torsion and metric-affine gravity.


\end{document}